\documentclass[manuscript,screen]{acmart}

\AtBeginDocument{%
	\providecommand\BibTeX{{%
			\normalfont B\kern-0.5em{\scshape i\kern-0.25em b}\kern-0.8em\TeX}}}

\usepackage{fancyhdr}
\usepackage[ruled, linesnumbered]{algorithm2e}
\usepackage{graphicx}
\usepackage{subfigure}
\usepackage{url}
\usepackage{multirow}
\usepackage{enumitem}
\usepackage{rotating}

\usepackage{amsmath,amssymb,amsthm}  
\usepackage{xcolor}
\usepackage{caption}

\usepackage{float} 

\newtheorem{define}{Definition}

\setcopyright{acmcopyright}
\copyrightyear{2022}
\acmYear{2022}
\acmDOI{XXXXXXX.XXXXXXX}

\setcopyright{none}
\renewcommand\footnotetextcopyrightpermission[1]{} 
\acmConference[XXX]{}{XXX}{}
\acmPrice{15.00}
\acmISBN{978-1-4503-XXXX-X/18/06}

\begin{document}

\title{Do as You Say: Consistency Detection of Data Practice in Program Code and Privacy Policy in Mini-App}

\author{Yin Wang}
\email{wy0724@stu.xjtu.edu.cn}
\author{Ming Fan}
\email{mingfan@mail.xjtu.edu.cn}
\author{Junfeng Liu}
\email{liujunfeng@stu.xjtu.edu.cn}
\author{Junjie Tao}
\email{taojunjie@stu.xjtu.edu.cn}
\author{Wuxia Jin}
\email{jinwuxia@mail.xjtu.edu.cn}

\affiliation{%
	\institution{Xi'an Jiaotong University}
	\city{Xi'an}
	\country{China}
}

\author{Qi Xiong}
\email{keonxiong@tencent.com}
\author{Yuhao Liu}
\email{drovliu@tencent.com}

\affiliation{%
	\institution{Tencent Cloud $\&$ Smart Industries Group}
	\city{Shenzhen}
	\country{China}
}

\author{Qinghua Zheng}
\email{qhzheng@mail.xjtu.edu.cn}
\author{Ting Liu}
\email{tingliu@mail.xjtu.edu.cn}
\affiliation{%
	\institution{Xi'an Jiaotong University}
	\city{Xi'an}
	\country{China}
}

\renewcommand{\shortauthors}{Yin Wang et al.}

\begin{abstract}

Mini-app is an emerging form of mobile application that combines web technology with native capabilities. Its features, e.g., no need to download and no installation, have made it popular rapidly. However, privacy issues that violate the laws or regulations are breeding in the swiftly expanding mini-app ecosystem. The consistency between what the mini-app does about the data in the program code and what it declares in its privacy policy description is important. But no work has systematically investigated the privacy problem of the mini-app before. In this paper, to our best knowledge, we are the first to conduct the compliance detection of data practice and policy description in mini-apps. 

To achieve this purpose, there are two main challenges. First, the mini-app is a novel form of application, and there lacks an effective tool to support us directly in obtaining the accurate data practice from code. Second, 
past consistency works have coarse-grained issues on data types, data operations, and consistency patterns. In this paper, we first customize a taint analysis method based on data entity dependency network to adapt to the characteristics of the JavaScript language in the mini-apps. Then, we transform data types and data operations to data practices in program codes and privacy policies, so as to finish a fine-grained consistency matching model.

We crawl 100,000 mini-apps on WeChat client in the wild and extract 2,998 with a privacy policy. Among them, only 318 meet the consistency requirements, 2,680 are inconsistent, and the proportion of inconsistencies is as high as 89.4$\%$. The inconsistency in the mini-app is very serious. Based on 6 real-world cases analyzed, in order to reduce this potential data leakage risk, we suggest that the developer should reduce the collection of irrelevant information and the straightforward use of templates, and the platform should provide data flow detection tools and privacy policy writing support. \footnote{New Paper}

\end{abstract}



\maketitle
\section{Introduction}\label{sec:introduction}

A new format of mobile application, called mini-app\cite{miniappwhitepaper} or host app, is getting popular nowadays\cite{ma2019app}. More and more consumers are using them for shopping, payments, education and travel. Merchants develop brand-specific mini-apps to sell goods, such as Bubble Mart. The mini-apps have similar functions to Android apps, but are more convenient than Android apps. They rely on Web technologies but also integrate with capabilities of the native apps. The size of mini-apps is limited by the host platform, generally smaller than 20MB, and their Web-like structure makes them free from installation. According to the statistical agency Aladdin Institute's article\cite{aldarticle} and Statista's report\cite{statisticsappnumber}, the number of WeChat mini-apps has achieved 3.8 million, which is even more than the Android apps on Google Play (3.5 million).

However, the data leak issues in these mini-apps have been easily neglected. Mini-apps could access many kinds of personal information such as name and gender\cite{ma2019appstorekiller}. A report\cite{nandureport} from the China Academy of Information and Communications found that among 52 commonly used mini-apps, 25$\%$ would share user data with third parties. Only 38.5$\%$ of mini-apps had a document describing information processing, which is often referred to as a privacy policy.

\begin{figure}
    \setlength{\abovecaptionskip}{5pt}
    \setlength{\belowcaptionskip}{3pt}
    \centering
    \subfigure[User interface and program code]{
        \includegraphics[width=0.68\textwidth]{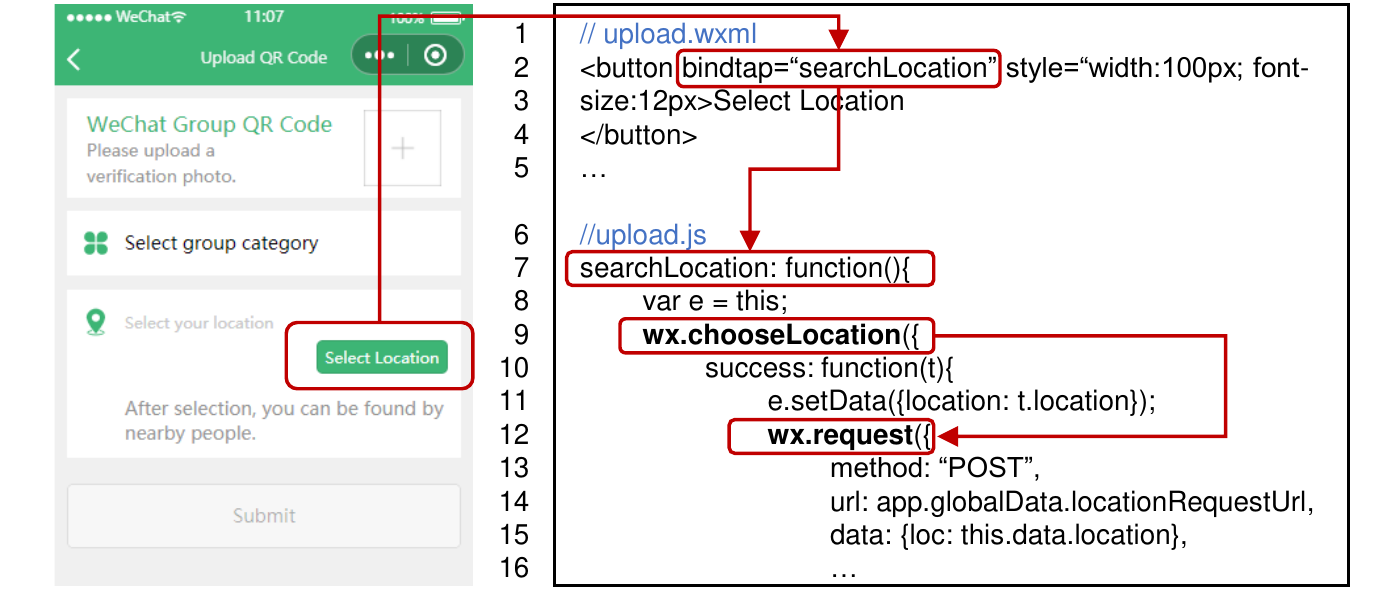}
    }
    \subfigure[Privacy policy]{
        \includegraphics[width=0.29\textwidth]{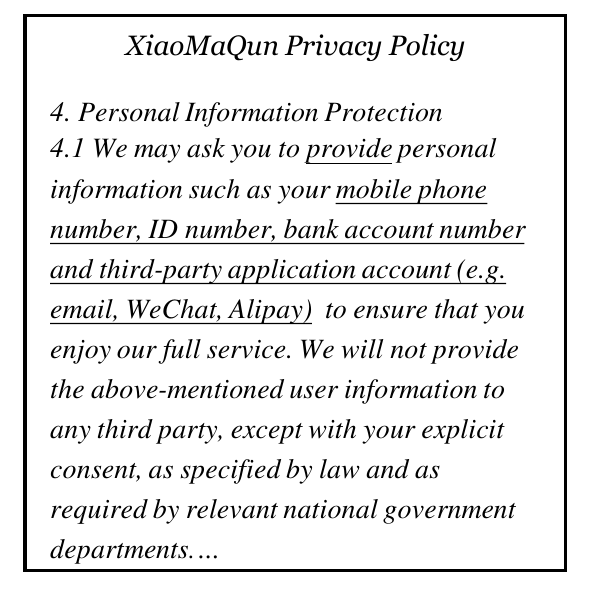}
    }
\caption{An example of a mini-app that says and does not agree. When the user clicks the \textit{"Select location"} button, his geographic location is returned via \textit{"wx.chooseLocation"} and then sent via \textit{"wx.request"}. However, this behavior is not mentioned in the privacy policy document it provides.}
\label{fig:sample}
\vspace{-20pt}
\end{figure}

Regulations related to personal information protection explicitly prohibit this behavior. China's latest mobile application personal information protection regulation\cite{chinesestandard} requires mobile applications to be open and transparent. It asks mobile applications should clearly inform users how to obtain and use data. The EU General Data Protection Regulation\cite{GDPR} also states that data processing should meet "lawfulness, fairness and transparency".

Unfortunately, most mini-apps do not meet the above requirements. For example, Figure \ref{fig:sample} shows a mini-app for sharing communication groups, called XiaoMaQun. On its share page, when the user clicks the \textit{"Select Location"} button, the \textit{"searchLocation"} function bound to the button is called. The user's geographical location is obtained by \textit{"wx.chooseLocation"} and finally is transferred to an external server by \textit{"wx.request"}. However, this behavior is not mentioned in the document it provides to explain the collection of information. The act of obtaining user data without notification violates the consistency of behavior and description.

In the past, problems such as inconsistency and noncompliance have been studied in Android apps. For different analysis objects and targets, relevant research can be divided into two categories: program behavior analysis and privacy policy analysis. The goal of program behavior analysis is to search the acquisition and transmission of sensitive information. Some studies focus on data passing in functional code\cite{enck2014taintdroid,xue2018ndroid,arzt2014flowdroid,octeau2013effective,li2015iccta,gordon2015information,pan2018flowcog,li2017static}, while others focus on data collection in user interface\cite{nan2015uipicker,huang2015supor,huang2016detecting,andow2017uiref,he2020textexerciser}. Privacy policy analysis uses natural language processing technology, focusing on the description of data collection and use, generally combined with program analysis work to detect contradiction and noncompliance issues in apps\cite{yu2018ppchecker,slavin2016toward,zimmeck2016automated,andow2020policheck,bui2021purliance,jin2018they,wang2018guileak,fan2020empirical}.

Disappointingly, the traditional detection method in Android apps is difficult to perform in the mini-app scenario. There are two main challenges as follows:

\textbf{Low availability of mini-app analytical tools.} Most of the research in the past was carried out around the Android apps, lacking the analysis tool of the mini-apps. The first and most critical difference between them is the language difference. Mini-apps use the JavaScript language as their logic language. Compared with the strongly typed language Java of Android apps, the dynamic characteristics\cite{chugh2009staged, hedin2012information} of JavaScript bring many obstacles to behavior analysis, such as the scope-based execution and flexible variable types. Second, most of the tools around JavaScript is focused on vulnerability checking and lack available interface for taint analysis, such as SAFE\cite{park2017analysis} and Understand\cite{Understand}, which makes it difficult to migrate and expand on mini-apps. Finally, there are some platform's predefined object and predefined APIs(\textit{e.g.}, asynchronous \textit{"wx.chooseLocation"} and the transfer function \textit{"setData"} in Figure \ref{fig:sample}). These characteristics can cause fractures in the data flow analysis results of traditional tools, such as TAJS\cite{jensen2011modeling}.

\textbf{Coarse granularity of consistency detection methods.} First, some past methods use rough data types in comparison, such as Polisis\cite{harkous2018polisis} only used 5 advanced data types and PolicyCheck\cite{andow2020policheck} only analyzed 12 data types supported by the data flow analysis tools AppCensus\cite{AppCensus}. Second, in terms of data operation analysis, some methods do not use data flow results to characterize behavior, which can lead to imprecision. For example, CHABADA\cite{gorla2014chabada} only extracts the frequency of API usage, and Purliance\cite{bui2021purliance} only extracts network traffic. Finally, in terms of the inconsistency patterns, in addition to the omitted disclosure and vague disclosure in Android apps\cite{andow2020policheck}, there is a new inconsistency in the mini-apps: the privacy policy will have serious redundant expressions that exceed the actual behavior. This issue has been ignored in previous studies.

In order to solve the above problems, we propose the following two methods:

\textbf{Taint analysis method based on data entity dependency network.} Aiming at the special language characteristics and engineering structure of mini-apps, we propose a taint analysis method based on data entity dependency network. We first parse the program code into an abstract syntax tree, then parse the code nodes associated with the data into entity units, establish entity dependencies according to the structure of the syntax tree, and finally perform taint analysis based on the relationship network.

\textbf{Fine-grained consistency matching method based on data practice model.} Aiming at the diverse data processing behaviors in mini-apps, we propose a data practice based comparison method. To achieve fine-grained behavior descriptions, data processing behaviors are split into combinations of data types and data operations. We first build two dictionaries of data types and operations. In addition, the APIs in the WeChat mini-app development manual are grouped into three types according to their functions. Program code behavior and privacy policy descriptions are transformed into data practice sets. Based on this model, we use five patterns (intersection, separation and three kinds of overlap) and two strengths (strong and weak) to study the inconsistency in mini-apps.

The main contributions of this work are: 

\begin{itemize}
    \item This work proposes a taint analysis method for the mini-app scenario for the first time. On a total of 620 data streams in 7 mini-apps, this method achieves an accuracy of 84.8$\%$ and can make up for 84.6$\%$ of the lack caused by asynchronous data flow and UI interaction events in existing tools.
    \item It is the first inconsistency analysis work carried out in the mini-app scenario. Experimental results show that that there are serious inconsistencies in 89.4$\%$ of the 2,998 mini-apps tested.
    \item  We implements a consistency detection system for WeChat mini-apps. About 100,000 mini-apps are collected and 2,998 privacy policies are extracted. To facilitate further research, system codes and datasets are available\cite{MCDS}.
\end{itemize}
 
\vspace{-5pt}
\section{Background}
\label{sec:background}

\subsection{WeChat Mini-app}

The first mini-app was released on the WeChat platform in 2017. Nowadays, WeChat has become the largest and most popular platform for mini-apps\cite{aldarticle}. There are more than 3 million mini-apps on the WeChat platform, and 450 million daily active users. Therefore, we choose the WeChat mini-app as our research object.

There are two main differences between mini-apps and traditional Android apps. First, mini-apps are developed by using the JavaScript dynamic programming language while apps are developed by using the Java, which is a static programming language. Second, mini-apps can only be accessed and run on its host platform while apps are executed in virtual machine. For example, the WeChat mini-apps need to run on the WeChat platform, i.e., the WeChat app. 

\subsection{Privacy Policy}

The privacy policy for applications is a legal document and software artifact to help users understand their information processing, such as what information and why the developers collect and use, how users can update, manage, export, and delete the information. 

The description related to information processing in the privacy policy is usually a combination of processing subject, data type, behavior and purpose. As shown in Figure 1, the statement on the right includes subject (\textit{"We"}), behavior (\textit{"ask you to provide"}), data type (\textit{"mobile phone number, ..."}), and purpose (\textit{"ensure that you enjoy our full service"}).

However, since the privacy policies are written in natural language without a uniform structure, it is difficult to directly obtain the four corresponding parts from the sentences. Furthermore, the data type or behavior that imply the same semantic meanings might be written using different words, such as the \textit{"user contact"} usually also refers to the user's \textit{"mobile phone number"} .

\subsection{Data Protection Regulations}

In recent years, different regions have successively promulgated regulations related to data protection. China's Personal Information Protection Law\cite{ChinaProtectionLaw} was released in 2021. It requires that the processing of personal information should follow the principles of openness and transparency, i.e., disclosing the personal information processing rules, including the purpose, method and data scope. European GDPR\cite{GDPR} (General Data Protection Regulation) ART 5 mentions that the data processing should meet the requirements of "lawfulness, fairness and transparency". In addition, the privacy rights mentioned by CCPA\cite{CCPA} include "the right to know about the personal information a business collects about them and how it is used and shared". The above laws require that data processing should allow users to know and agree. When the documents provided by the application do not match the actual behavior of the application, the precondition of the right to know is denied.

Therefore, in this work we propose a consistent comparison method between the actual behavior and the description about data processing in the mini-app scenario. Through program analysis and description analysis, it can help users and developers to understand whether the program meets the transparency principles proposed by the regulations.

\vspace{-5pt}
\section{Approach}
\label{sec:approach}

\begin{figure}[t]
    \centering
    \resizebox{\linewidth}{!}{
        \includegraphics{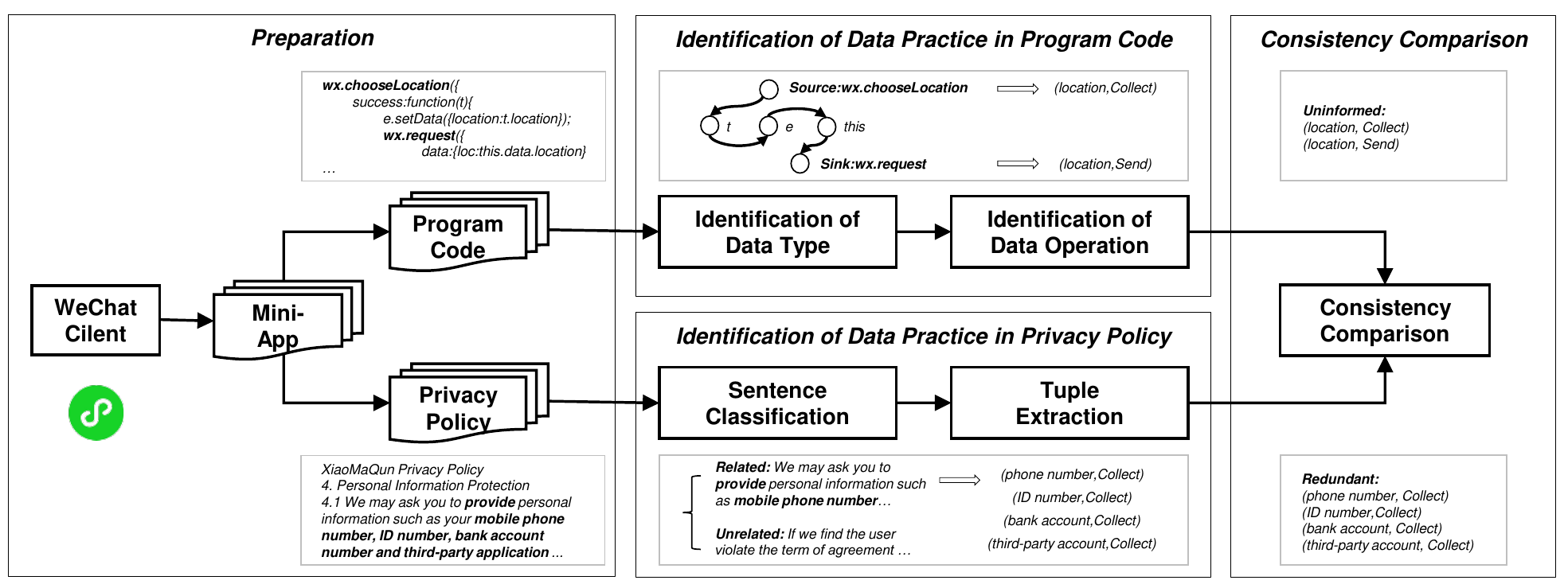}
    }
    \caption{The overview architecture of our approach, which contains four procedures, including the preparation, the identification of data practice in program code and privacy policy, and the consistency comparison. }
    \label{fig:arch}
    \vspace{-15pt}
\end{figure}

The overview architecture of our approach is illustrated in Figure \ref{fig:arch}, which contains four procedure: 1) In the preparation procedure we crawl a set of mini-apps and extract their program codes and privacy policies; 2) In the first identification procedure, we identify the data practice in the program code using static taint analysis method; 3) In the second identification procedure, we extract the sentence description as tuples from the privacy polices using natural language analysis method;  4) In the consistency comparison procedure, we compared the data practices of the two to check for inconsistencies.

\subsection{Preparation}

In the preparation stage, there are three steps, including the mini-app downloading, the extraction of program codes and the obtaining of privacy policies.

As for where to download the mini-apps, unlike the Google Play for the Android app users, there is no open market for the mini-app users. Therefore, to obtain a set of mini-apps, we first  crawl a name list of mini-apps from several third-party websites which make recommendations. Note that these sites only provide the mini-apps' name or QR code to enter the mini-apps, not the code package. Then, we expand the list based on the search results on the WeChat client. For example, when searching for "Meituan Takeaway", the list of search results will not only give the mini-app with the same name, but also the mini-apps with extended names, such as "Meituan Order", "Meituan Smart Hardware" and "Mimi Takeaway".

After downloading the mini-app set, considering that their corresponding code packages are usually compressed, we use an open source unpacking tool called WxAppUnpacker\cite{wxappUnpacker} to obtain the source code. The source code contain a main entry file \textit{app.js}, multiple page files, resource files, additional plugins and third-party library files. For each page, it contains a logic file in JavaScript format, layout file in WXML(WeXin Markup Language) format, style file in WXSS (WeXin Style Sheets) format and configuration file in Json format.

There are two main presentation modes of the privacy policies, one is attaching the text to the code statically, the other one is dynamic linking, which will only be displayed when the specified page is open. For the code attaching mode, we directly extract them from the static code by using a set of keywords such as \textit{"privacy policy"} and \textit{"agreement"}. For the dynamic linking mode, we design a UI testing tool to search the policy text. This tool runs on the WeChat client through the automated testing framework Airtest\cite{Airtest}. Specifically, when it enters an mini-app, it first clicks the index button at the bottom to switch pages, and then clicks the text related to the privacy policy within each page. In addition to this, it also tries to explore on some intermediate pages, such as login registration and settings pages.

\subsection{Definition Of Data Practice}

In this section, we first give the definition of data practice as below.

\begin{define}[Data practice]
    Data practice is the behavior of data processing by the application. It is denoted as a tuple, (\textit{t,a}), where \textit{t} represents the data type, and \textit{a} represents the operations of data processing. 
\end{define}

For example, the data practices related to location in Figure \ref{fig:sample} can be represented by (\textit{location}, \textit{Collect}) and (\textit{location}, \textit{Send}). To accurately describe data practices, we constrain the scope of data types and data operations as follows.

\subsubsection{Data Type Categories} We rely on a national standard called GB/T 35273-2020\cite{chinesestandard} for mobile applications data collection and divide the data types into 13 primary categories and 80 secondary categories, which are listed in Table \ref{tab:typedict}. And each secondary category contains a set of phrases that present similar semantic meanings due to the language diversity. For example, the "\textit{address}" secondary category has the entities "\textit{city}" and "\textit{township}", which denote the address information in different ways.

Specifically, to expand the phrases under each secondary category, we first extract all the phrases from the UI layout files and privacy policies of 100 popular mini-apps in different categories. Then, we leverage the KNN algorithm to perform a phrase clustering, where the similarity calculation is performed based on the Word2vec\cite{mikolov2013word2vec} algorithm. Finally, we manually check and revise the phrases set for each secondary category to ensure its accuracy.     

\begin{table}[tb]
    \setlength{\abovecaptionskip}{5pt}
    \setlength{\belowcaptionskip}{0pt}
    \caption{Privacy-related Entity Dictionary (Data Type)}
    \label{tab:typedict}
    \centering
    \begin{footnotesize}
    \setlength{\tabcolsep}{4mm}{
    \begin{tabular}{ccccc}
    \toprule
    Primary Category & Secondary Category & Num & Phrase & Num \\ \midrule
    Basic & name, birthday, gender, address, ... & 11 & sex, city, mobile, ... & 352 \\
    Identify & ID card, passport, driver's license, ... & 8 & identity card, visa, ... & 72 \\
    Biometric & recognition features & 1 & touchId, fingerprint ... & 20 \\
    Network & account, password, photo, video, ... & 14 & pwd, uid, album, ... & 271 \\
    Health & weight, medication records, ... & 6 & hospital, doctor, ... & 62 \\
    Work$\&$Education & occupation, salary, organization, ... & 5 & job, company, graduate ... & 108 \\
    Property & bank account, credit, transaction, ... & 6 & order, income, payment, ... & 183 \\
    Communication & sms, address book, friends list, ... & 5 & chatList, smsVerify, session, ... & 41 \\
    Web Log & log, browsing records, click record, ... & 5 & search history, favorites list, ... & 140 \\
    Device & phone model, config, app & 3 & iOS, android, IMEI, sysinfo, ... & 78 \\
    Location & position, accommodation & 2 & gps, geographical location, ... & 99 \\
    Hardware & camera, recorder, bluetooth, ... & 11 & wifi, nfc, ram, ... & 68 \\
    Other & marriage, express, travel & 3 & destination, departure date, ... & 53 \\ \hline
    Total &  & 80 &  & 1547 \\ \bottomrule
    \end{tabular}%
    }
    \end{footnotesize}
    \vspace{-5pt}
    \end{table}

    \begin{table}[tb]
        \setlength{\abovecaptionskip}{5pt}
        \setlength{\belowcaptionskip}{0pt}
        \caption{Privacy-related Entity Dictionary (Data Operation)}
        \label{tab:operationdict}
        \centering
        \begin{footnotesize}
        \setlength{\tabcolsep}{6mm}{
        \begin{tabular}{ccc}
        \toprule
        Operation Category & Operation Words & Num \\
        \midrule
        Collect & provide, acquire, read, collect, input, access, record, accept, shoot, scan, ... & 33 \\
        Use & process, display, store, pay, analyze, call, save, show, search, dail, ... & 42 \\
        Send & share, pass, transmit, upload, transfer, post, submit, spread, receive & 9 \\
        \bottomrule
        \end{tabular}
        }
        \end{footnotesize}
        \vspace{-5pt}
        \end{table}

\subsubsection{Data Operation Categories} Data operations are grouped into three categories based on different processes of data processing, including \textit{Collect}, \textit{Use}, and \textit{Send}. Here, \textit{Collect} refers to acquiring data, \textit{Use} refers to processing data within the program, and \textit{Send} refers to sending data from the  inside to the outside. Table \ref{tab:operationdict} lists our constructed vocabulary for three operation categories and their corresponding set of operation words.

\subsubsection{Data Practice Examples} 
\begin{figure}[t]
    \centering
    \resizebox{\linewidth}{!}{
        \includegraphics{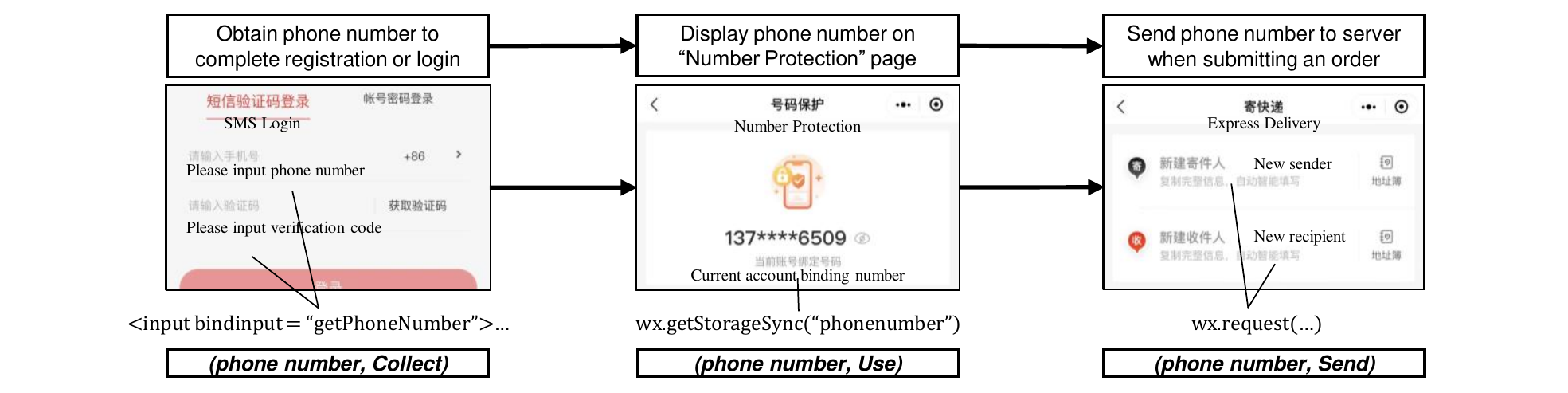}
    }
    \caption{Data practice example from a courier mini-app.}
    \label{fig:datapractice}
    \vspace{-10pt}
\end{figure}

Figure \ref{fig:datapractice} illustrates three data practice examples of the same data type \textit{"phone number"} and the three different data operations. First, the users are required to enter the mobile phone number when registering and logging in, which constitutes a tuple of (\textit{phone number, Collect}). Second, this number is read from the cache and displayed on a "Number Protection" page, which constitutes a tuple of (\textit{phone number, Use}). Finally, when the user fills in the recipient and sender information to submit an order, the number will be sent to the server, which constitutes a tuple of (\textit{phone number, Send}).

\subsection{Identification of Data Practice in Program Code}

The identification of data practice in code relies on the static data flow analysis method, which is divided into two steps: identification of data type and identification of data operation. 

\subsubsection{Identification of Data Type} Data types will be specified according to different data sources:

\begin{itemize}
    \item \textbf{Platform-provided API.} \textit{t} is determined by the data content of the return value. For example, the data content returned by \textit{"wx.chooseLocation"} is geographic location information, so \textit{t = location}.
    \item  \textbf{User Input.} \textit{t} is determined by the text of the input associated component. In Figure \ref{fig:datapractice}, the prompt text of the input component \textit{"Please input phone number"} contains the keyword \textit{"phone number"}, so \textit{t = phone number}.
\end{itemize}

The first rule relies on the type division of the APIs, however, to the best of our knowledge, there is currently no research on the taint analysis of mini-apps for reference. Some related studies only focus on the usage of a specific API\cite{wang2022characterizing}, which is incomplete for our work. In order to solve this problem, we carefully read the API list of the developer manual provided by the WeChat, and divided these APIs into different categories according to their return content. As listed in Table \ref{tab:sourceandsink}, there are 150 source APIs such as \textit{"wx.getSystemInfo"} that returns the system information and \textit{"wx.chooseImage"} that accesses albums. 

\begin{table}[t]
    \setlength{\abovecaptionskip}{5pt}
    \setlength{\belowcaptionskip}{0pt}
    \caption{Source APIs provided by WeChat platform}
    \label{tab:sourceandsink}
    \begin{footnotesize}
        \setlength{\tabcolsep}{5mm}{
        \begin{tabular}{ccc}
            \toprule
            Category  & Example & Num \\ 
            \midrule
            System  & wx.getSystemInfo, wx.getSystemInfoSync, env & 3 \\
            Storage & wx.getStorage, wx.getStorageSync, wx.getStorageInfo, wx.getStorageInfoSync & 4 \\
            Network  & wx.getNetworkType, wx.request, wx.downloadFile, wx.onSocketMessage & 5 \\
            Media  & wx.getImageInfo, wx.chooseImage, wx.getVideoInfo, wx.openVideoEditor & 26  \\
            Location  & wx.chooseLocation, wx.getLocation, wx.onLocationChange & 3   \\
            File  & wx.getSavedFileList, wx.getFileInfo, wx.getSavedFileInfo & 13  \\
            Share  & wx.getShareInfo, wx.authPrivateMessage & 2  \\
            Device & wx.getBeacons, wx.onBeaconSeviceChange, wx.onBLEPeripheralConnectionStateChanged & 76   \\
            Open Interface & wx.login, wx.getAccountInfoSync, wx.addCard  & 16   \\
            Lifecycle & wx.getEnterOptionsSync, wx.getLaunchOptionSync & 2  \\ 
            \midrule
            Total & & 150  \\
            \bottomrule
            \end{tabular}
        }
        \end{footnotesize}
        \vspace{-5pt}
\end{table}

\begin{table}[t]
    \setlength{\abovecaptionskip}{5pt}
    \setlength{\belowcaptionskip}{0pt}
    \caption{Sink APIs provided by WeChat platform}
    \label{tab:sink}
        \begin{footnotesize}
        \setlength{\tabcolsep}{6mm}{
        \begin{tabular}{ccc}
            \toprule
            Category  & Example & Num \\ 
            \midrule
            Usage  & wx.setStorage, wx.saveFile, wx.showToast, wx.saveImageToPhotosAlbum, console.log, ... & 120 \\
            Transmission & wx.request, wx.uploadFile, wx.requestPayment, wx.connectSocket, wx.sendSocketMessage, ...  & 31 \\
            \bottomrule
            \end{tabular}
            }
        \end{footnotesize}
        \vspace{-15pt}
\end{table}

The second rule requires extracting the information provided by the user interface from the layout file. The relevant components contain the event-triggering component and the form submission component. In the case of Figure \ref{fig:sample}, the semantics about the data type can be directly parsed according to the description text (e.g. \textit{"Select location"}) or the bind event function name (e.g. \textit{"searchLocation"}). This part is appended to the text properties of the related component.

However, we find that some related components have no text description themselves. To solve the problem, we utilize the work of UIPicker\cite{nan2015uipicker}. Specifically, nodes in the layout file will be resolved into a tree of components according to their containment relationships. For the missing description component, it searches the associated node text from the bottom up, and each time a node is accessed, it will search that node and all the node's sub-node text until the associated description is found.

\subsubsection{Identification of Data Operation} For a data practice of a certain type \textit{t}, its data operations \textit{o} are specified according to the following rules:
\begin{itemize}
    \item \textbf{Collect.} When there is a source of \textit{t}-type data (calling the API provided by the platform or user input), the data practice can be represented by (\textit{t, Collect}).
    \item \textbf{Use.} When the data flow of \textit{t}-type data passes through a usage type sink API (e.g. \textit{"wx.setStorageSync"} for storage), the data practice can be represented by (\textit{t, Use}).
    \item \textbf{Send.} When the data flow of \textit{t}-type data passes through a transmission category sink API (e.g. \textit{"wx.request"} for sending a request), the data practice can be represented by (\textit{t, Send}).
\end{itemize}

Among the above three operations, \textit{Send} and \textit{Use} depend on the type of sink API at the end of the data flow. To distinguish different endpoints, the APIs that accept data are divided into two categories: usage and transmission, which is shown in Table \ref{tab:sink}. The division between them depends on whether the data is sent outside in their official description.

\subsubsection{Data Flow Analyze}
The analysis of operation depends on the results of data flow analysis. We first parse program code into the dependency network on different code entities and then perform data flow analysis based on the network.

\begin{define}[Code entitiy]
    Code entity is a general term for elements in code, including variables, functions, classes, and files, etc., and can be used to build program dependency graphs.
\end{define}

Here we choose the three entities most closely related to the data dependency graph as the analysis targets, including variables, files and functions. The dependency resolution procedures are shown in Algorithm \ref{alg:entitychain} and \ref{alg:dependency}.

\begin{algorithm}[t]
    \caption{Build Entity And Scope Chain}
    \label{alg:entitychain}
    \footnotesize
    \KwIn{Mini-app program code $C$, Empty scope chain  $S= \varnothing$.}
    \KwOut{Scope chain  $S$.}
    \SetKwFunction{ParseToAbstractSyntaxTree}{ParseToAbstractSyntaxTree}
    \SetKwFunction{GenerateNewEntity}{GenerateNewEntity}
    \SetKwFunction{BuildEntity}{BuildEntity}
    \SetKwFunction{addEntity}{addEntity}
    \SetKwFunction{GetScope}{GetScope}
    \SetKwFunction{GenerateNewScope}{GenerateNewScope}
    \SetKwFunction{BuildScope}{BuildScope}
    \SetKwFunction{isSubScopeOf}{isSubScopeOf}
    \SetKwFunction{addScope}{addScope}
    \SetKwFunction{buildScopeChain}{buildScopeChain}
    \BlankLine

    $T$ = \ParseToAbstractSyntaxTree($C$); \tcp*[h]{get the abstract tree of program code C}

    $S$.\addScope($T$);

    \For{each $n$ in $T$}
    {
        \If(\tcp*[h]{an entity generation node(variables, functions and files)}){\GenerateNewEntity($n$)}{
            $newEntity$ = \BuildEntity($n$);

            $nearestScope$ = \GetScope($n$, $S$); \tcp*[h]{get the current and nearest scope of $n$ from $S$}

            $nearestScope$.\addEntity($newEntity$); \tcp*[h]{each entity belongs directly to the nearest scope}
        }

        \If(\tcp*[h]{a scope generation node}){\GenerateNewScope($n$)}{
            $newScope$ = \BuildScope($n$)\;
            \For{each $m$ in $S$}{
                \If{$newScope$.\isSubScopeOf($m$)}{
                    $m$.\buildScopeChain($newScope$); \tcp*[h]{child scope $newScope$ is linked with parent scope $m$}
                }
            }
            $S$.\addScope($newScope$)\;
        }

    }
\end{algorithm}

\begin{algorithm}[t]
    \caption{Build Dependency}
    \label{alg:dependency}
    \footnotesize
    \KwIn{Scope chain  $S$.}
    \KwOut{Data Dependency Graph $DDG$.}
    \SetKwFunction{getEntity}{getEntity}
    \SetKwFunction{BuildSetDependency}{BuildSetDependency}
    \SetKwFunction{BuildUseDependency}{BuildUseDependency}
    \BlankLine

    \For{each $m$ in $S$}{
        \For{each $n$ in $m$.abstractSyntaxTreeNodeList}{
                $relatedEntities$ = \getEntity($n$, $m$);

                \If{$n.type$ in [AssignmentExpression(left value), CallExpression(setData), Property, ReturnStatement, ...]}{
                    \tcp*[h]{add "Set" dependencies for entities that change entity values}

                    $DDG$.\BuildSetDependency($relatedEntities$);
                } 

                \ElseIf{$n.type$ in [AssignmentExpression(right value), CallExpression(except setData), LogicalExpression, ...]}{
                    \tcp*[h]{add "Use" dependencies for entities that reference entity values}
                    
                    $DDG$.\BuildUseDependency($relatedEntities$);
                }
            }
    }
\end{algorithm}

(1) We use a NodeJs library called \textit{esprima} to parse the mini-app code into an abstract syntax tree, and determine the target entities according to the node type through predefined rules. \textit{"GenerateNewEntity"} in Algorithm \ref{alg:entitychain} is used to determine whether a node will generate an entity object, such as \textit{"var a = 1"} will generate a variable entity \textit{"a"}. 

(2) According to the grammar rules of JavaScript and WeChat mini-app, some nodes may generate different code scopes that affect data dependency graph construction. We create corresponding scope objects and add the above entities to these objects, and connect these scopes in a chain to solve it. \textit {"buildScopeChain"} in Algorithm \ref{alg:entitychain} will link the scopes with the containment relationship, and the child scope can access the objects under the parent scope.

(3) As the Algorithm \ref{alg:dependency} is shown,the nodes will be traversed in each scope and build dependencies for the aforementioned entities based on whether the nodes involve data transfer or not. Nodes that can generate a transfer relationship include assignment statements, function parameter transfer, return values, and \textit{"setData"} defined by the WeChat mini-app. For example, \textit{"a.setData(b)"} will assign \textit{"b"} to \textit{"a.data"}.

\begin{figure}
    \vspace{-18pt}
    \subfigure[Program code]{
        \includegraphics[width=0.35\linewidth]{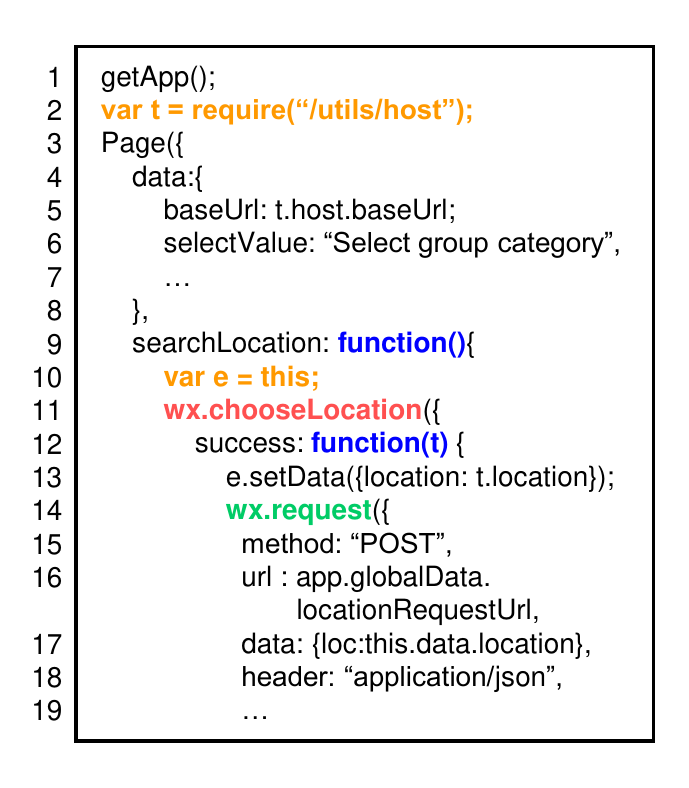}
    }
    \subfigure[Dependency resolution and flow search on abstract tree]{
        \includegraphics[width=0.5\linewidth]{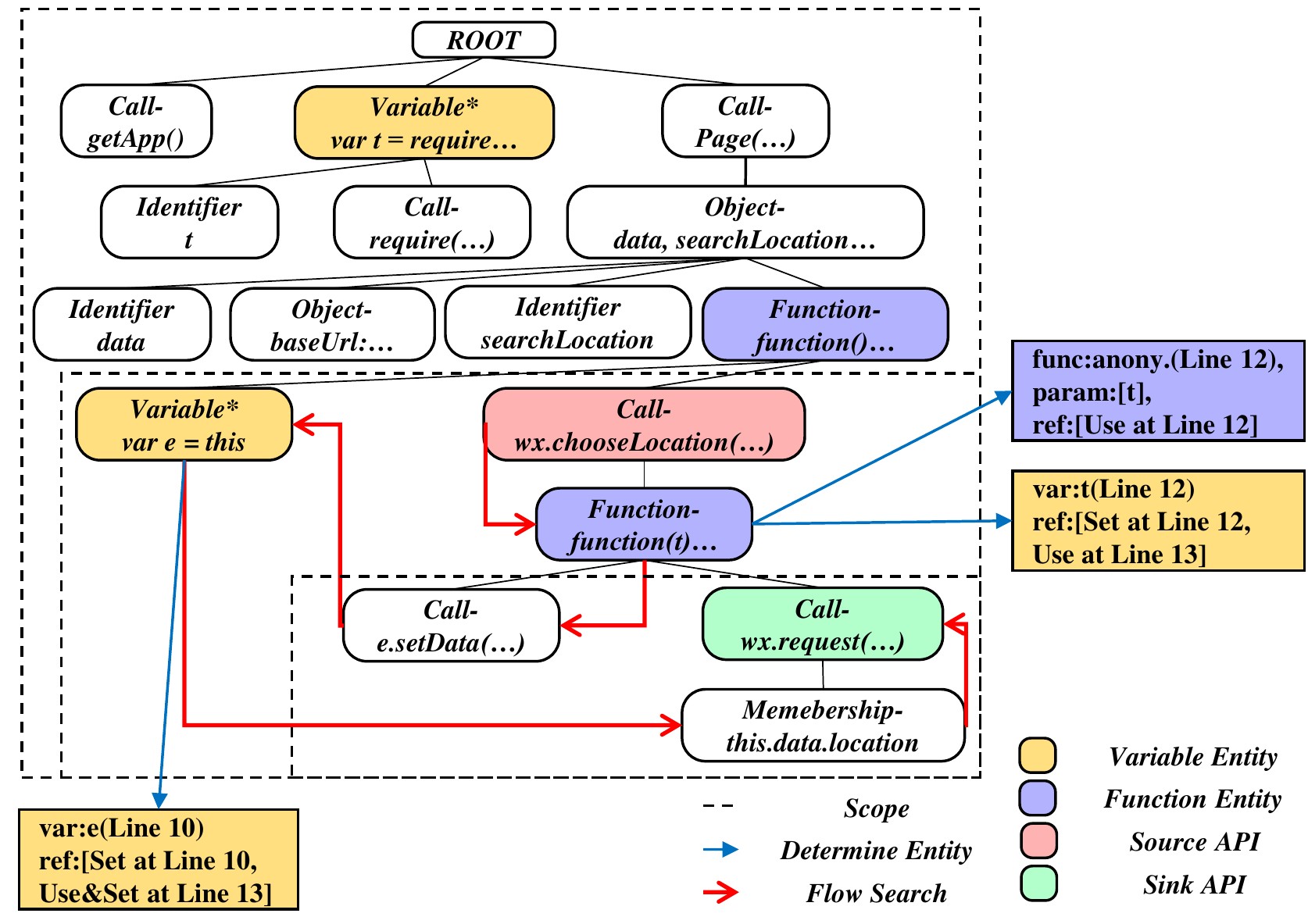}
    }
    \caption{Examples of dependency resolution and flow search. The left side is an extension of the sample code in the introduction section, and the right side is the abstract syntax tree of the code.}
    \label{fig:dependencyflow}
    \vspace{-20pt}
\end{figure}

Taking Figure \ref{fig:dependencyflow} as an example, the code on the left will first be parsed into the abstract syntax tree. First, the entire tree is traversed to find the node that generates the entity, including variable declaration statements (yellow nodes) on lines 2 and 10, function parameter on line 12, and anonymous functions declaration on lines 9 and 12 (blue nodes).

Then, as divided by the dotted line diagram on the tree, the code will be divided into three levels of scope. The division of scope is to determine the correct dependencies of variables. For example, the variable \textit{"t"} is defined in lines 2 and 12, and the variable \textit{"t"} is used in lines 5 and 13. Based on the scope division, \textit{"t"} in line 5 will form a dependency with \textit{"t"} in line 2, and \textit{t} on line 13 will form a dependency with \textit{"t"} on line 12.

Finally, we look for nodes with transitive behavior and add \textit{"Set"} dependencies to them. In the example, the variable definitions in lines 2 and 10, the callback function parameters in line 12, the predefined function \textit{"setData()"} in line 13, and the assignment of properties of objects in lines 4, 13, and 17 are regarded as data transfer. Their dependencies are constructed as a connection of related entities.

After construction, data flow search is a reachable search process from the starting point to the end point on the network. A complete flow is shown by the red line in the Figure \ref{fig:dependencyflow}. The source API \textit{"wx.chooseLocation"} first returns location information to \textit{"t"} (line 12), \textit{"t"} has transitive dependencies through \textit{"setData"} and \textit{"e"} (line 10), and finally passes it to a sink API \textit{"wx.request"} through the \textit{"this"} keyword. This path constitutes the data practice of (\textit{location, Send}).

\subsection{Identification of Data Practice in Privacy Policy}

We use the natural language processing method to determine the data practice in privacy policy.  Considering that the types and operations in the statement are expressed in the form of entities and relationships, we determine the data practice according to the following rules:

\begin{itemize}
    \item \textbf{Data type.} For the entities in the statement related to data processing, if there is an entity description similar (e.g. \textit{"city"}) to the data type we specified (e.g. \textit{"address"}), \textit{t} is the corresponding type category.
    \item \textbf{Data operation.} For a \textit{t}-type entity, if the predicate (e.g. \textit{"provide"}) of \textit{t} is within the three data operation types defined and \textit{o} is the data operation corresponding to the predicate, this data practice is represented by \textit{(t, o)}.
\end{itemize}

The entities are meaningful things in a sentence and the predicates describe the connections between entities. The identification is divided into two steps: sentence classification and tuple extraction.

\subsubsection{Sentence Classification} A privacy policy generally includes data irrelevant sentences. There would be a lot of unrelated sentences in the extraction procedure. In order to precisely locate and extract the descriptive text, we first split the privacy policy into separate sentences, and then classify them into data related and data unrelated sentences. Specifically, we construct a dataset of different sentences. One paper participant and three trained students participated in the labeling of the dataset. By labeling 50 privacy policies, 2136 sentences related to data processing were marked. 

After that, we leverage the bag-of-words model for sentence classification. Each sentence is converted into a 1547-dimensional vector (consistent with the vocabulary dimension), with each dimension indicating whether the sentence contains the word in the corresponding position, and fed into different trained classifiers. For example, when using \textit{"mobile number", "bank account", "location"} as the vocabulary, the privacy policy snippet in Figure \ref{fig:sample} is transformed into a (1,1,0) vector. The models are trained using different classification algorithms, including MLP, CNN and LSTM. The detailed parameters and classification effects of different models will be introduced in the Section \ref{sec:evaluation}.  

\subsubsection{Tuple Extraction} After the sentence classification, we use word similarity calculation method to extract key information from the sentence. First the sentence is split into tuples of entities and relationships using a triple extraction tool LTP\cite{che2020n}, then we use data types vocabulary for entities and data operations vocabulary for relationships to extract similar words respectively. 

For example, the sentence in Figure \ref{fig:sample} will be parsed into four tuples including \textit{(we, ask, you), (you, provide, personal information such as ...), (you, ensure, you enjoy our full service)} and \textit{(you, enjoy, our full service)}. Then, entity words like \textit{"personal information like ..."} and \textit{"our full service"} will be compared with type words, and relationship words like \textit{"provide"} and \textit{"ensure"} will be compared with operation words. Text similarity methods include overlap coefficient, cosine distance, Euclidean distance, and Dice coefficient. Words that meet the similarity threshold will be extracted. For example, the former (data type) will get \textit{"mobile phone number"}, \textit{"ID number"}, \textit{"bank account number"} and \textit{"third-party account number"}, and the latter (data operation) will be extracted to get \textit{"provide"}.

\subsection{Consistency Comparison} 

After completing the analysis of data practices in the program code and the privacy policy, the process of consistency comparison can be regarded as a comparison between the two data practice sets. 

\begin{figure}
    \setlength{\abovecaptionskip}{5pt}
    \setlength{\belowcaptionskip}{10pt}
    \subfigure[Intersection]{
        \includegraphics[width=0.16\linewidth]{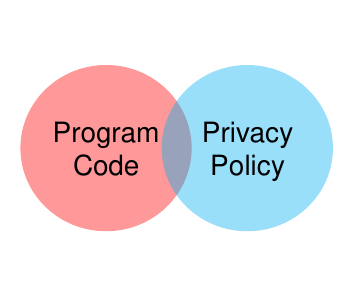}
        \label{fig:consistencymode-intersect}
    }
    \subfigure[Separation]{
        \includegraphics[width=0.16\linewidth]{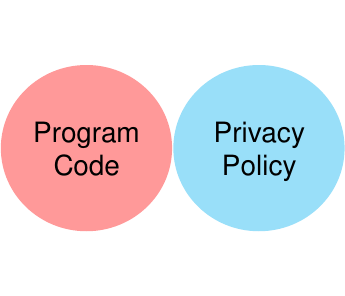}
    }
    \subfigure[Overlap(Uninformed)]{
        \includegraphics[width=0.16\linewidth]{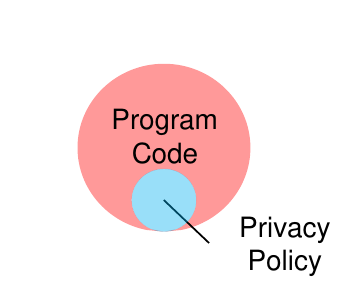}
    }
    \subfigure[Overlap(Redundant)]{
        \includegraphics[width=0.16\linewidth]{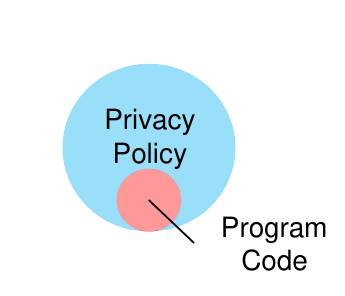}
    }
    \subfigure[Overlap(Consistent)]{
        \includegraphics[width=0.16\linewidth]{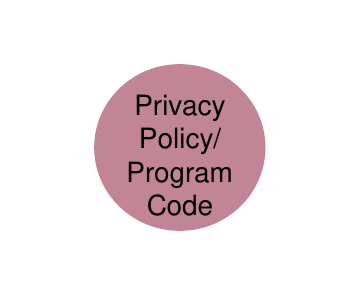}
    }
    \caption{Five patterns of consistency: intersection, separation, and three types of overlap. The non-overlapping parts of the first four subgraphs are inconsistencies. "Uninformed" means that some data practices appear in the program code, but do not appear in the privacy policy. "Redundant" means that some data practices appear in the privacy policy, but do not appear in the program code.}
    \label{fig:consistencymode}
    \vspace{-20pt}
\end{figure}

As shown in Figure \ref{fig:consistencymode}, the relationships of sets can be divided into five patterns: intersection, separation, and three types of overlap. Among the three types of overlap, we use "Uninformed" to represent that some data practices appear in the program code but do not appear in the privacy policy, and use "Redundant" to represent that some data practices appear in the privacy policy, but do not appear in the program code. The non-overlapping parts of the first four subgraphs are inconsistencies and the last one statisfies the consistency requirements.

Taking into account the different severities of inconsistencies, the comparison results are divided into weak inconsistencies and strong inconsistencies. Weak inconsistency occurs when a data type results on both sides but the data operations are different. Strong inconsistency occurs when the two results have a completely different data types. 

\begin{figure}
    \setlength{\abovecaptionskip}{5pt}
    \setlength{\belowcaptionskip}{5pt}
    \includegraphics[width=0.9\linewidth]{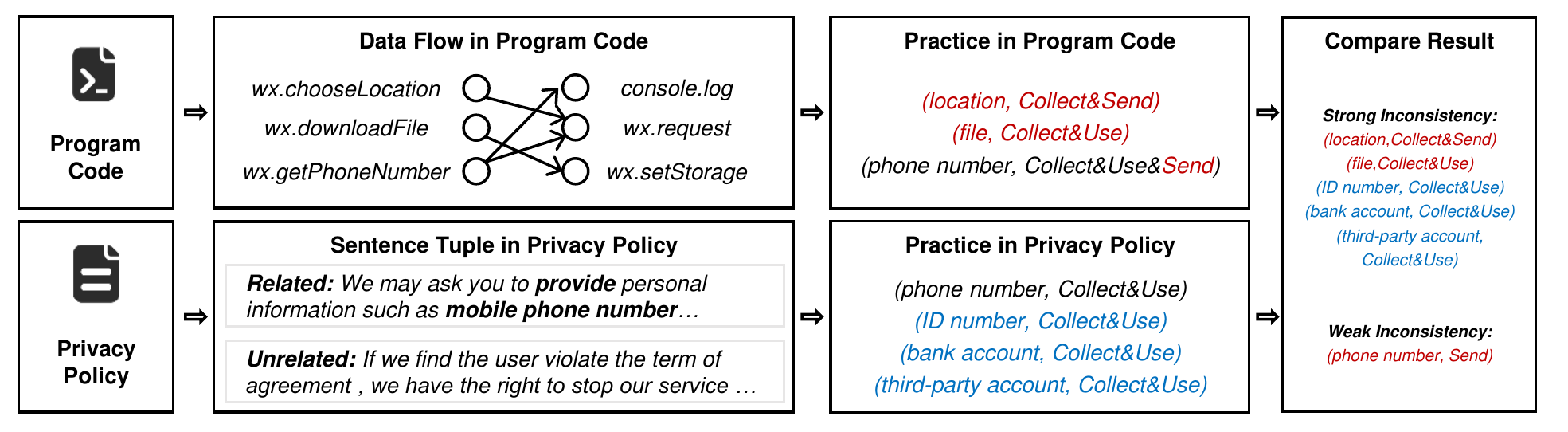}
    \caption{An example of consistency comparison. According to the severity of the inconsistenies, the results were divided into strong inconsistency and weak inconsistency.}
    \label{fig:consistency}
    \vspace{-20pt}
\end{figure}

In the example shown in Figure \ref{fig:consistency}, we obtains the connection network between source and sink APIs through data flow analysis on program code, which is transformed into 7 data practice tuples, and extracts the associated statements through the statement classification method on privacy policy, which are transformed into 8 data practice tuples. The comparison results belong to the intersection class shown in Figure \ref{fig:consistencymode-intersect}. Strong inconsistencies exist in \textit{"location"}, \textit{"document"} in the program code and \textit{"ID number"}, \textit{"bank account"}, \textit{"third-party account"} in the privacy policy. Weak inconsistency exists in the \textit{"phone number"} mentioned in both results, and the \textit{"Send"} behavior in the program code is not mentioned in the privacy policy.

\vspace{-5pt}
\section{Evaluation}
\label{sec:evaluation}

\subsection{Research Question}

We conducted our experiments based on the following five questions.

\textbf{RQ1: How accurate is our method in extracting data practice from the mini-app's program code?}

\textbf{RQ2: What is the distribution of data practice in the mini-apps' program codes?}

\textbf{RQ3: How accurate is our method in extracting data practice from the mini-app's privacy policy?}

\textbf{RQ4: What is the distribution of data practice in the mini-app's privacy policies?}

\textbf{RQ5: What is the inconsistencies distribution in mini-apps? How serious are these inconsistencies?}

\subsection{Exprimental Setup}

\begin{figure}[t]
    \setlength{\abovecaptionskip}{5pt}
    \setlength{\belowcaptionskip}{0pt}
    \subfigure[Quantity distribution]{
        \includegraphics[width = 0.40\linewidth]{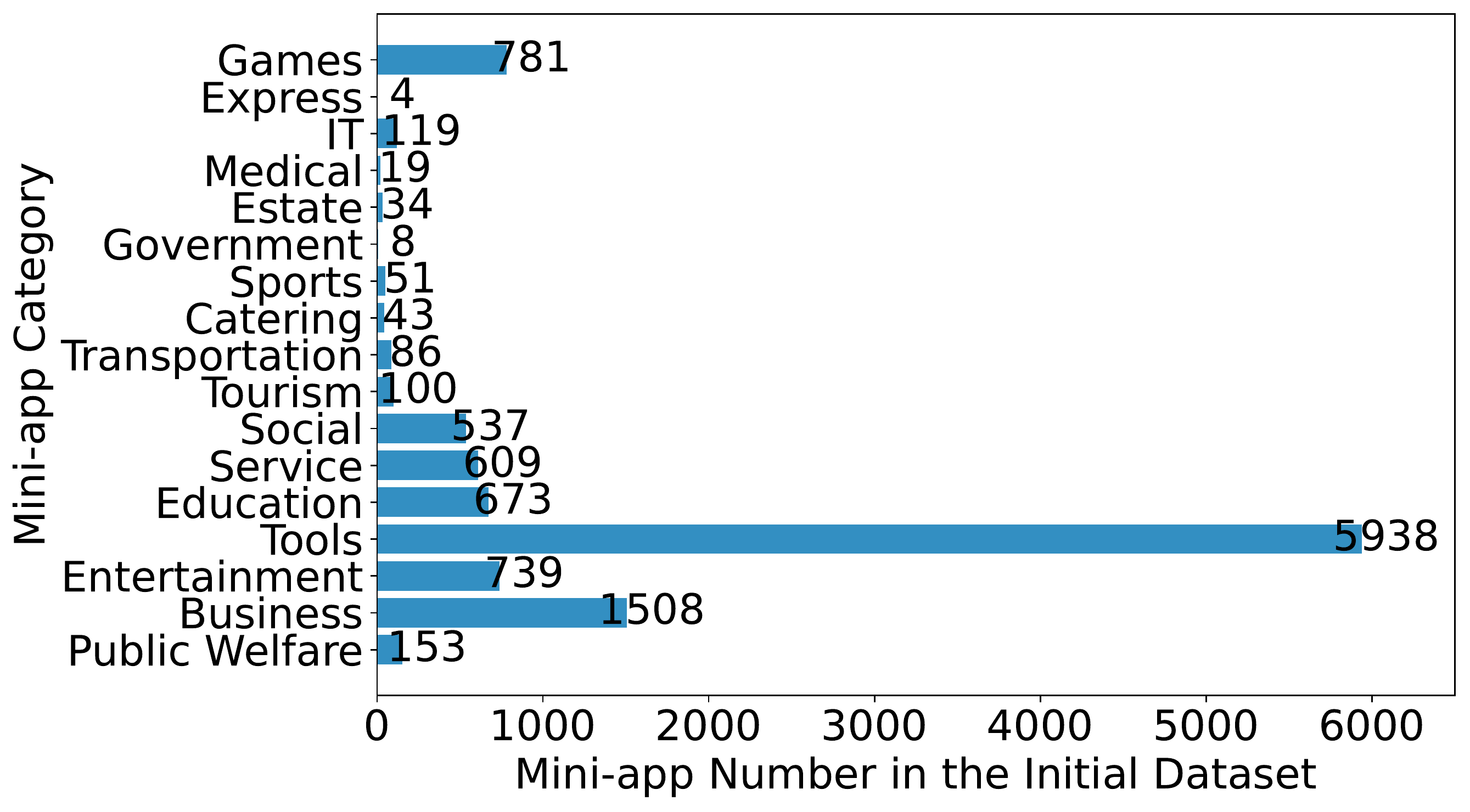}    
    }
    \subfigure[Download time distribution]{
        \includegraphics[width = 0.40\linewidth]{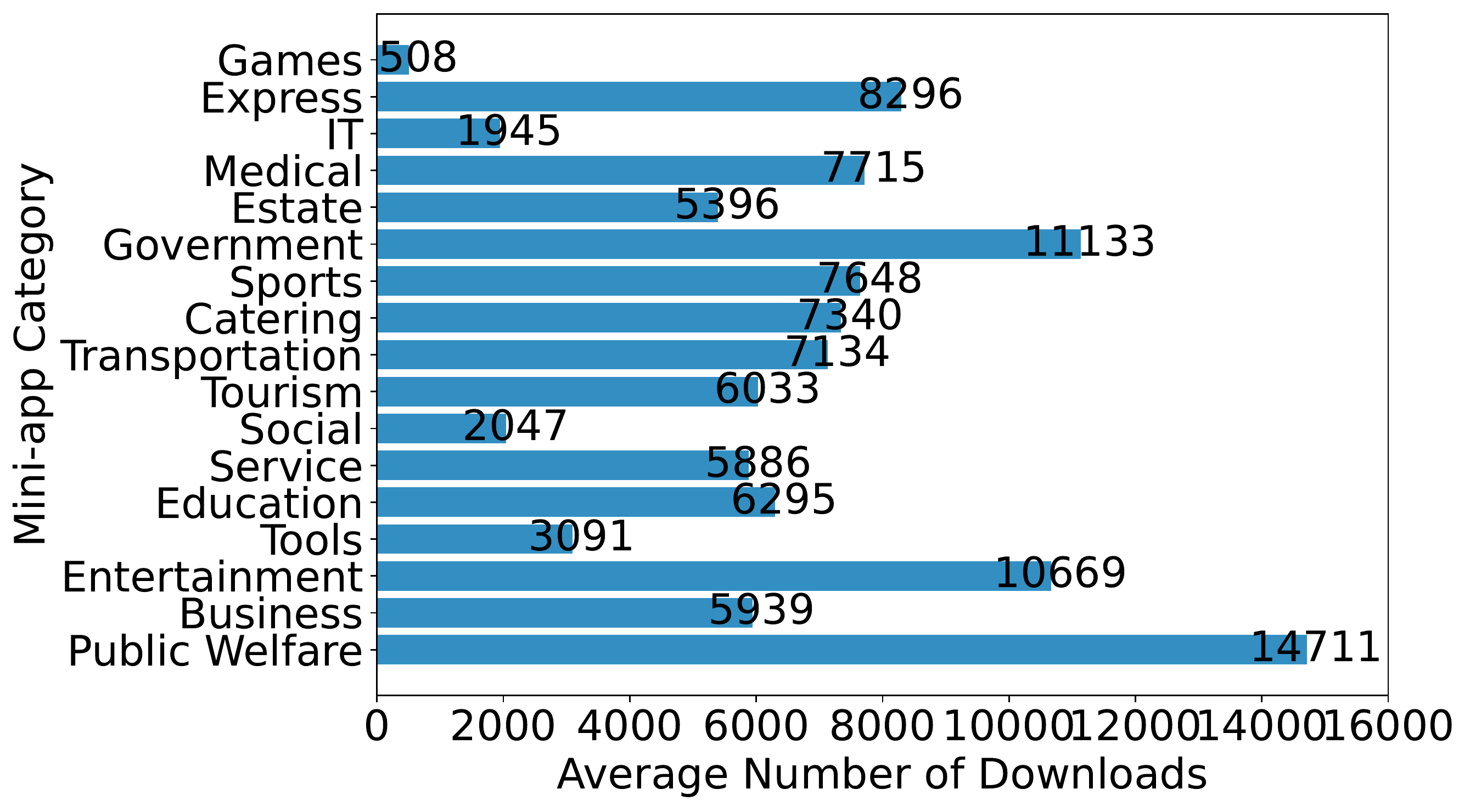}    
    }
    \caption{Initial dataset distribution. This part contains 12,000 mini-apps, which are the basis for subsequent search and crawling.}
    \label{fig:datasetdistribution}
    \vspace{-20pt}
\end{figure}

\textbf{Dataset.} We first obtained a list of 12,000 mini-apps from third-party websites. As shown in Figure \ref{fig:datasetdistribution}, the average downloads of different categories of mini-apps ranged from 508 to 14,711. Among them, the number of tool-type mini-apps is the largest, at 5,938, and the number of express, medical, government, and catering mini-programs is less than 100. Then we expand the search on the client-side based on this list. Since the code package obtained by crawling does not carry downloads and categories information, the distribution of the final data set is difficult to obtain. We finally crawled about 100,000 mini-apps code packages and extracted 2,998 mini-apps with privacy policies as our base data set. There were some privacy polices we didn't find, because for dynamically attached privacy policies, the testing tool cannot guarantee full access to them.

\textbf{Metrics.} We use classical accuracy and callback rates as the main evaluation metrics. Accuracy refers to the proportion of correct results in all samples, and this value is used to measure whether the method can give correct results. Recall rate refers to the proportion of all true values in the correct results, which is used to measure omissions in results.

\textbf{Environment.} We accomplished experiments on a server with three Intel Xeon Gold 6266C CPU, one NIVDIA GeForce RTX 3090 GPU and one 15TB hard disk. The operating system version is Ubuntu 18.04. 

\subsection{Accuracy of Taint Analysis Method}

Since there is no relevant benchmark, we manually marked 7 randomly selected samples as shown in Table \ref{tab:flowvalidity}. We first divided data flows into three categories according to their sources: APIs called synchronously, APIs called asynchronously, and user interface inputs. 

Specifically, taking Figure \ref{fig:sample} as an example, \textit{"setData"} in line 6 is a synchronous API, and the statement in this line will be executed immediately in the body of the function declared in line 5. The \textit{"wx.chooseLocation"} in line 11 is an asynchronous API. After the user selects the location successfully, the location data will be passed to the parameters of the callback function. The \textit{"searchLocation"} in line 2 is a user-triggered event, and the event is an anonymous function in line 2. If it is an input type component, the data entered by the user will be passed to the parameter list of the anonymous function after the event is triggered.

The difference between mini-apps and traditional JavaScript taint analysis lies in the latter two types of data flow. For some JavaScript static analysis tools such as TAJS\cite{jensen2011modeling}, the data flow caused by the synchronous API can be analyzed given the function name of the data source. However, the asynchronous API customized by the WeChat platform and the data transfer caused by the unique UI trigger events cannot be obtained. As shown in Table \ref{tab:flowvalidity}, the data streams of three parts are 95, 328 and 197 respectively, accounting for 15.3$\%$, 52.9$\%$ and 31.7$\%$. It can be seen that the data flow analysis method that does not adapt to the mini-apps will lose 84.6$\%$ of the data flow analysis results.

During testing we also found that there were some broken data flows. Of the 620 calibrated data streams, 524 are expected, 94 are broken and no result is wrong. The main reason for the fracture is that we can't solve the built-in objects provided by JavaScript very well. For example, in \textit{"a.push(b)"}, when \textit{"a"} is an array type, \textit{"push"} will construct a data dependency for \textit{"b"} and \textit{"a"}. A simple solution is to fix the built-in property list. However, \textit{"a"} may also be a custom object and \textit{"push"} is a custom function and does not transfer data. We need to know the exact variable type, but this is very difficult for dynamically typed languages.

\noindent\fbox{\parbox[c]{0.95\linewidth}{\textit{\textbf{Answer to RQ1:} The data flow analysis method we propose makes up for the loss of 84.6$\%$ of other existing analysis tools in the mini-app scenario, and 84.8$\%$ of all results are in line with expectations. This method can effectively analyze the data processing behavior from the mini-app code.}}}

\begin{table}[t]
    \setlength{\abovecaptionskip}{5pt}
    \setlength{\belowcaptionskip}{0pt}
    \caption{Taint analysis results, including data flow distribution (based on source) and manual inspection results.}
    \centering
    \footnotesize
    \begin{tabular}{cccccccc}
    \toprule
    \multirow{2}{*}{Mini-App Name} & \multicolumn{4}{c}{Flow Distribution (based on source)} & \multicolumn{3}{c}{Inspection} \\
    & Sync. API & Async. API & UI & All & Right & Broken & Wrong \\ \midrule
    998 Excavator  & 3 & 192 & 77 & 272 & 260 & 12 & 0 \\
    CMC Theater  & 2 & 38 & 7 & 47 & 46 & 1 & 0 \\
    Hi Repair  & 28 & 52 & 16 & 96 & 55 & 42 & 0 \\
    BaiKa Butler & 0 & 6  & 3 & 9 & 7 & 2 & 0 \\
    Animated Universe & 52 & 18 & 60 & 130  & 92 & 38 & 0 \\
    Registration Assistant$*$ & 1 & 2  & 0 & 3 & 3 & 0 & 0 \\
    Hand Over Cards$*$ & 9  & 20 & 34 & 63 & 63 & 0 & 0 \\ \hline
    Total & 95 & 328 & 197 & \textbf{620}  & \textbf{526} & \textbf{94} & \textbf{0} \\ \bottomrule
    \multicolumn{8}{l}{$*$ \footnotesize{These codes are obfuscated and the abstract syntax tree structure is broken.}}
    \end{tabular}%
    \label{tab:flowvalidity}
    \vspace{-20pt}
    \end{table}

\subsection{Data Practice Distribution in Program Code} 
\begin{figure}
    \setlength{\abovecaptionskip}{5pt}
    \setlength{\belowcaptionskip}{0pt}
    \begin{minipage}{0.7\linewidth}
        \subfigure[Source API Distribution]{
            \includegraphics[width=0.48\linewidth]{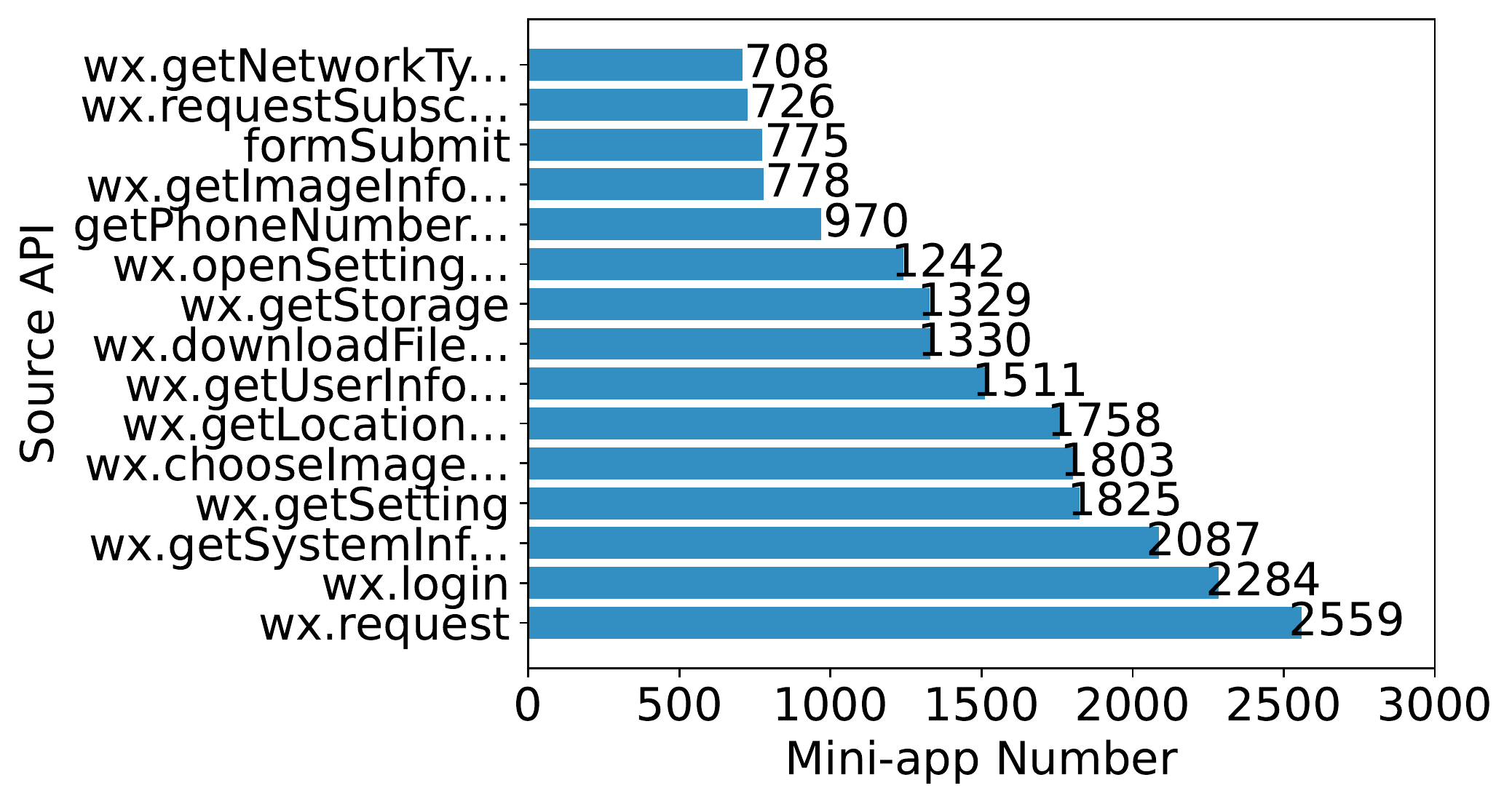}
            \label{fig:practicedistribution-a}
        }
        \subfigure[Sink API Distribution]{
            \includegraphics[width=0.48\linewidth]{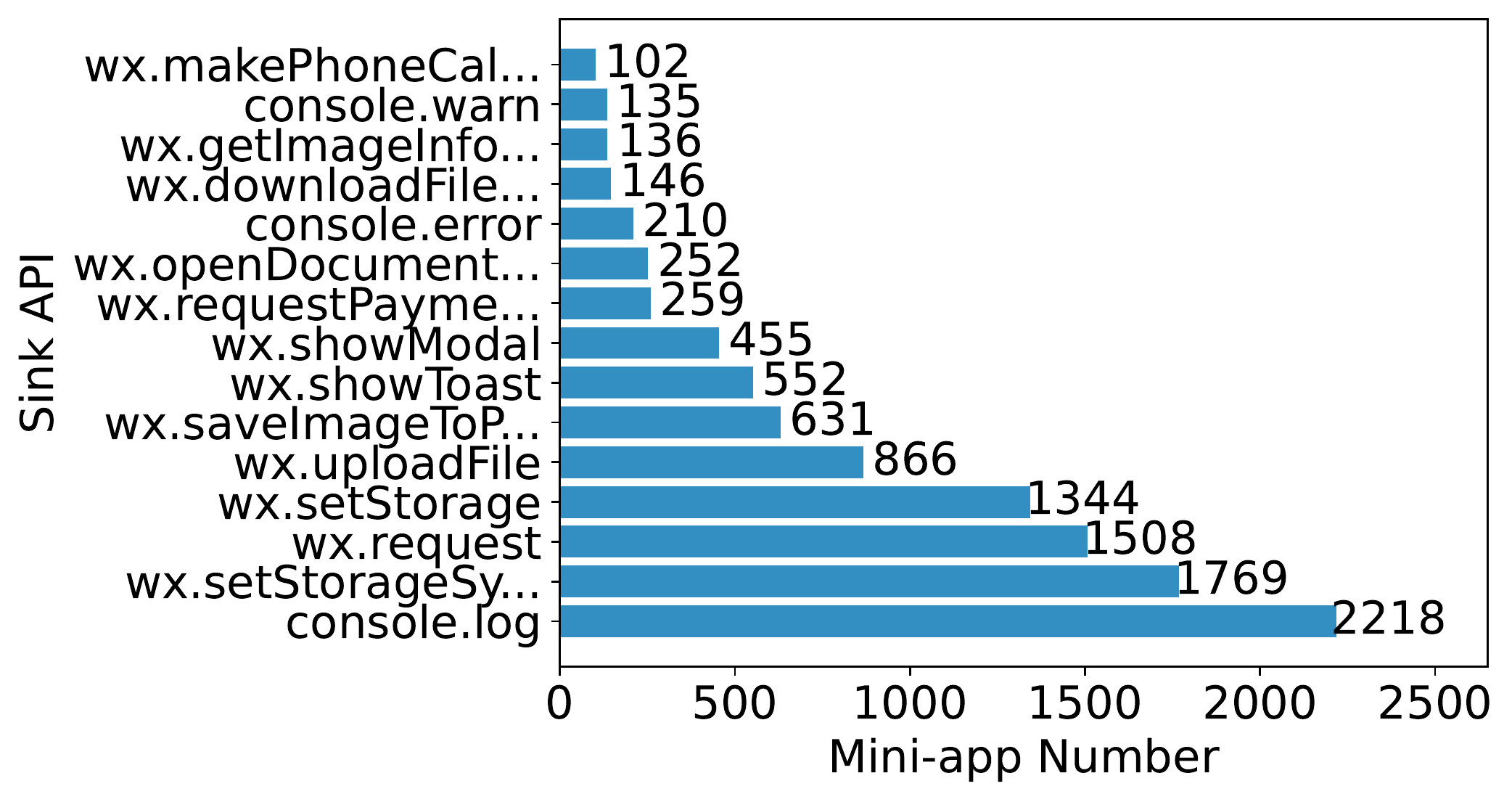}
            \label{fig:practicedistribution-b}
        }
        \subfigure[Data Type Distribution]{
            \includegraphics[width=0.48\linewidth]{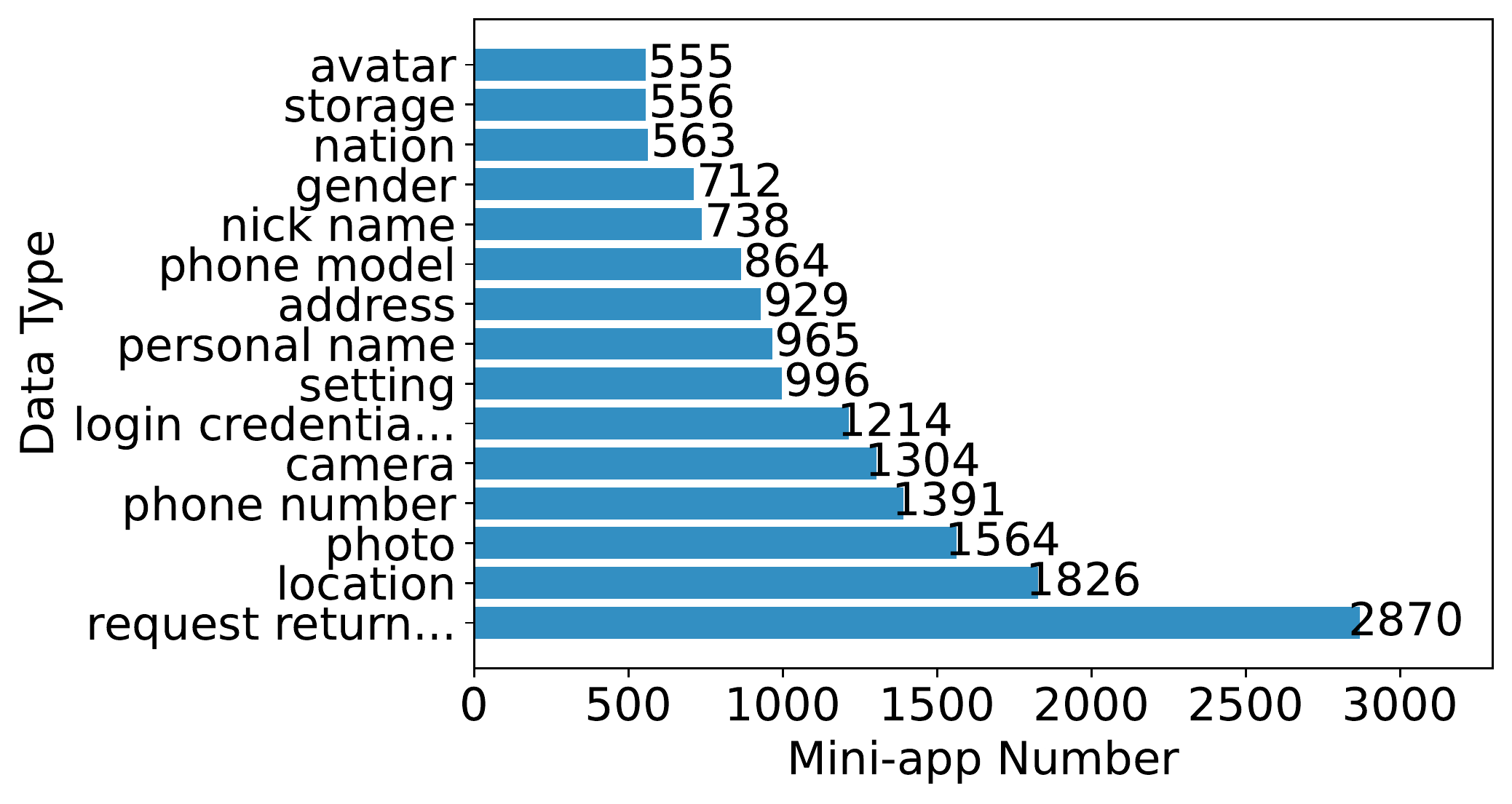}
            \label{fig:practicedistribution-c}
        }
        \subfigure[Data Operation Distribution]{
            \includegraphics[width=0.48\linewidth]{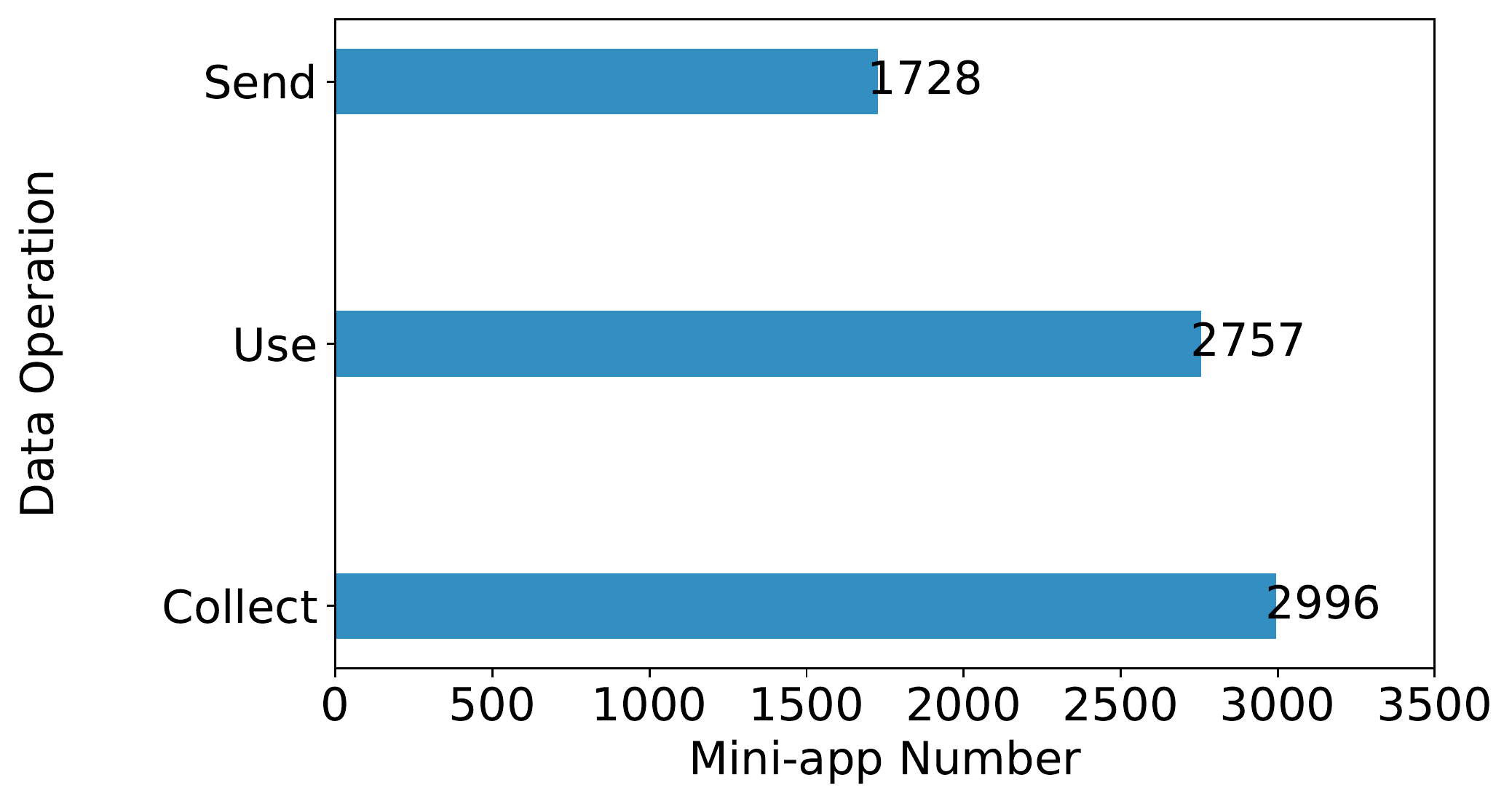}
            \label{fig:practicedistribution-d}
        }
    \end{minipage}
   \begin{minipage}{0.23\linewidth}
        \centering
        \subfigure[Data Practice Heatmap]{
            \includegraphics[width=\linewidth]{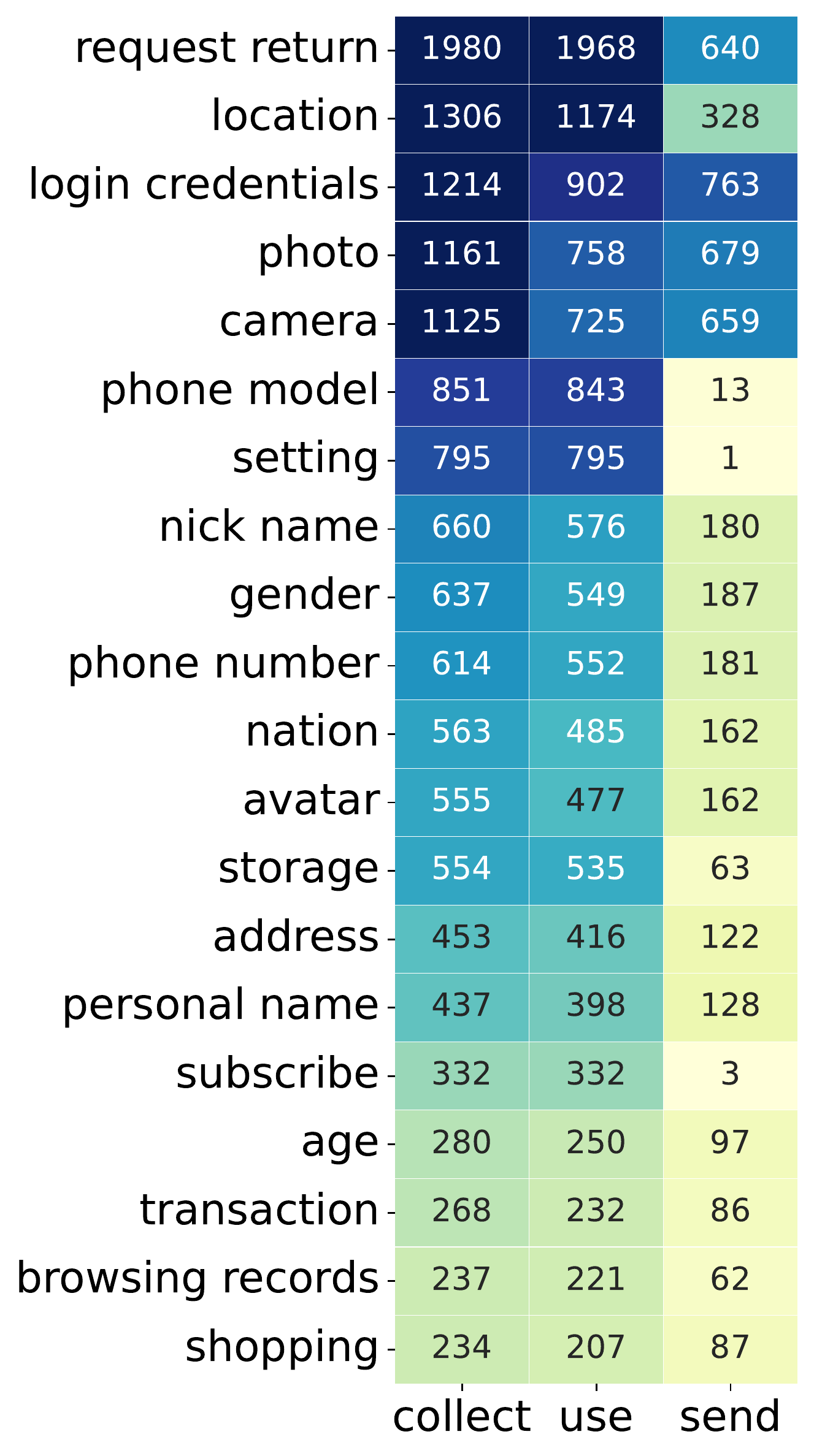}
            \label{fig:practicedistribution-e}
        }
   \end{minipage}
    \caption{Data practice distribution in program code. (a) and (b) represent the statistical results of data flow analysis, and (c) and (d) are the result of converting (a) and (b) to data practice attributes. (e) is the distribution of data practice in the code.}
    \label{fig:practicedistribution}
    \vspace{-5pt}
\end{figure}

The analysis result on 2,998 mini-apps' codes are shown in Figure \ref{fig:practicedistribution}. Figure \ref{fig:practicedistribution-a} and Figure \ref{fig:practicedistribution-b} are the distributions of different independent source and sink APIs. The results were transformed into the form of data practice attributes which is shown in Figure \ref{fig:practicedistribution-c} and Figure \ref{fig:practicedistribution-d}. Finally, we selected the top 20 data types and drew the distribution heat map of data practices, as shown in Figure \ref{fig:practicedistribution-e}. Through statistics, we have the following findings:

(1)The most frequently used sources come from the API interface provided by the WeChat platform. In Figure \ref{fig:practicedistribution-a}, more than 2,000 mini-apps use \textit{"wx.request"} for sending requests and \textit{"wx.login"} for user login. 

(2) Some custom event functions that get specific types of data are also very common. For example, more than 700 mini-apps customize \textit{"getPhoneNumber"} to obtain the user's mobile phone number and \textit{"formSubmit"} to submit form information. Developers have a common habit of naming functions.

(3) The three operations are very common in mini-apps. Figure \ref{fig:practicedistribution-d} also shows that 2,996(99.93$\%$)  mini-apps will collect, 2,757(91.96$\%$) will use and 1,728(57.64$\%$) will send user data. The proportion of usage behavior and collection behavior is close and the transmission behavior is more than half.

(4) The collection of user sensitive information is widespread. Figure \ref{fig:practicedistribution-e} shows that the most data practices focus on the communication data, such as request return (95.7$\%$) and login credentials (40.5$\%$). However, there are still more than 1,161 (38.73$\%$) mini-apps that collect users' sensitive location information and photo albums. 

\noindent\fbox{\parbox[c]{0.95\linewidth}{
\textit{\textbf{Answer to RQ2:} For data types, mini-apps tend to collect communication data (e.g. request returns (95.7$\%$) and login credentials (40.5$\%$)), as well as user sensitive information (e.g. location (61.2$\%$) and album(52.2$\%$)). For data operations, most of the data will be used within the mini-apps and external transmission is also very common in mini-apps. The three data operations collect, use and send account for 99.93$\%$, 91.96$\%$ and 57.64$\%$ respectively.}
}}

\subsection{Accuracy of Description Extraction Method} 

\subsubsection{Classifier Evaluation} We used different vector transformation methods and different classification models to find the best classification method. First, we picked 50 privacy policies, completed the annotation of sentences, and divided them into relevant sentences and irrelevant sentences. There were 2,136 sentences divided into related sentences. Then, we used a bag-of-words model to convert sentences into different vectors for different vocabularies. Finally, we input these vectors into the model to complete training and testing.

As shown in Table \ref{tab:sentenceclassify}, there are three vocabularies in the transform. "Type" means the original data type vocabulary, "Type$\&$Operation" means adding an operation vocabulary to the type vocabulary, "Type*Operation" means combining the type word and operation word into a bigram, and then forming a new vocabulary. Three models includes MLP, CNN and LSTM. In terms of parameter settings, MLP uses a 512-dimensional dense layer, CNN uses a three-layer convolutional network layer with dimensions of 256, 128 and 64, and LSTM uses a 256-dimensional lstm layer. 

Table \ref{tab:sentenceclassify} shows that when using the merged vocabulary and the MLP model, the classification of sentences is the best, with an accuracy of 97.98$\%$ and a recall rate of 92.56$\%$. We think there are two reasons:

(1) In terms of vocabulary selection, the target sentence usually contains data type related word and operator, but not always both. The extended vocabulary can extract more information from the sentence, but the premise of the extension is that the newly added words have a wide distribution in the sentence.

(2) In terms of model selection, the CNN model is generally used for image recognition and do not perform well for one-dimensional data, and the LSTM model requires more time-series features to be extracted from sentences. The aforementioned vector transformation methods cannot work well with these models.

\begin{minipage}[t]{\linewidth}
    \vspace{-25pt}
    \begin{minipage}[t]{0.5\linewidth}
        \centering
        \begin{table}[H]
            \setlength{\abovecaptionskip}{3pt}
            \setlength{\belowcaptionskip}{0pt}
            \caption{Sentences classification result}
            \centering
            \resizebox{0.9\textwidth}{!}{
            \footnotesize
            \begin{tabular}{ccccc}
            \toprule
            One-hot Vocabulary & Model & Accuracy & Recall & F1 Score \\ \midrule
            \multirow{3}{*}{Type} & MLP & 0.9578 & 0.7944 & 0.8669 \\
             & CNN & 0.9506 & 0.8043 & 0.8492 \\
             & LSTM & 0.8815 & 0.4971 & 0.5920 \\ \hline
            \multirow{3}{*}{Type\&Operation} & MLP & \textbf{0.9798} & \textbf{0.9256} & \textbf{0.9408} \\
             & CNN & 0.9725 & 0.9021 & 0.9189 \\
             & LSTM & 0.8936 & 0.4990 & 0.6187 \\ \hline
            \multirow{3}{*}{Type*Operation} & MLP & 0.9695 & 0.8412 & 0.9053 \\
             & CNN & 0.8598 & 0.1896 & 0.3188 \\
             & LSTM & 0.9037 & 0.5861 & 0.6780 \\ \bottomrule
            \end{tabular}
            }
            \label{tab:sentenceclassify}
            \end{table}
    \end{minipage}
    \begin{minipage}[t]{0.45\linewidth}
        \begin{figure}[H]
                \centering
                \setlength{\abovecaptionskip}{3pt}
                \setlength{\belowcaptionskip}{0pt}
                \includegraphics[width=0.6\textwidth]{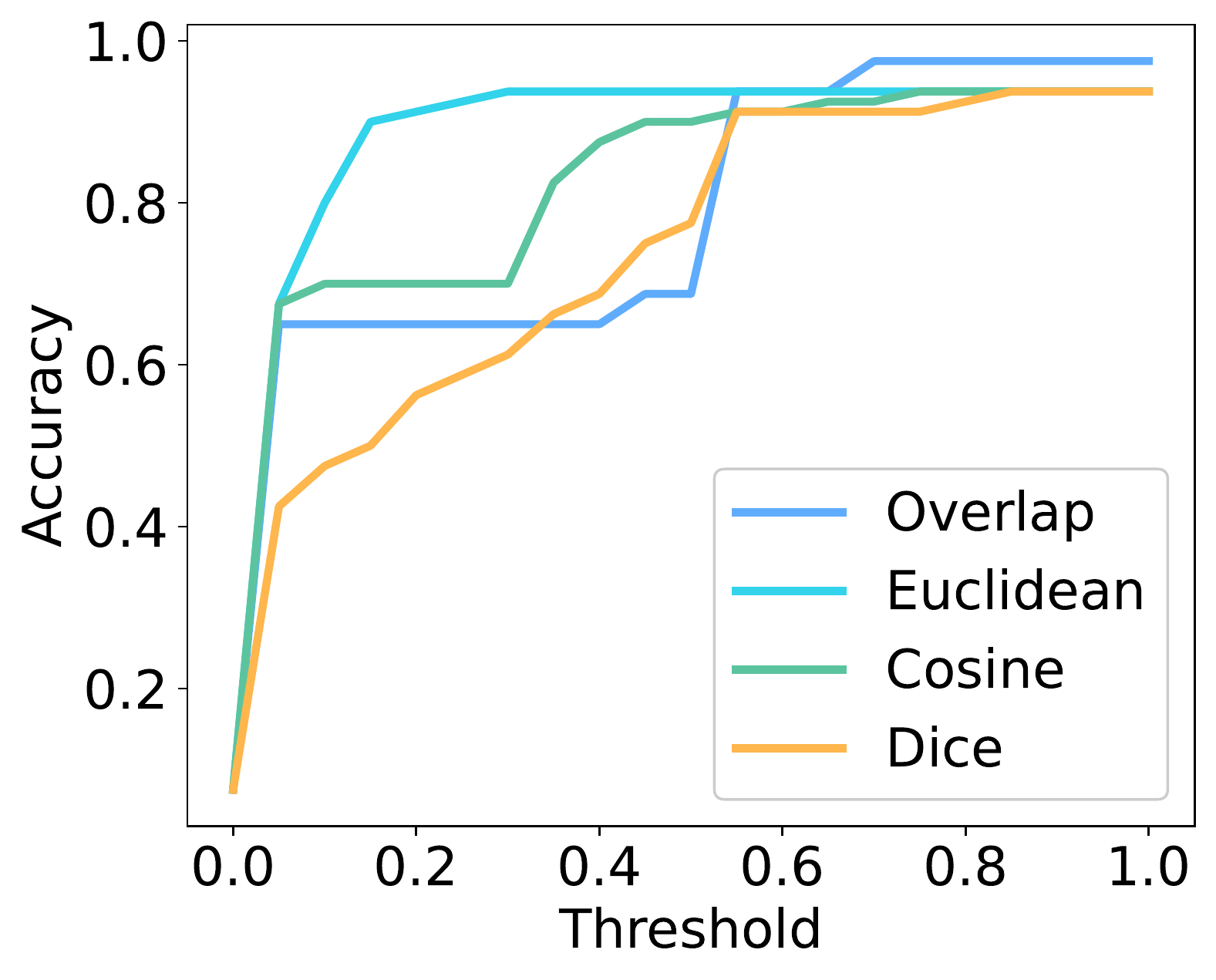}
                \caption{Effects of using different extraction methods}
                \label{fig:extraction}
            \end{figure}
    \end{minipage}
    \vspace{5pt}
\end{minipage}

\subsubsection{Tuple Extraction Evaluation} Here we discuss the effects of four similarity extraction methods. First, we randomly selected 306 relevant sentences from 2,136 and completed the annotation of data practice. Then, we used the LTP tool to extract the entities and relationships in the sentences, calculated the word similarity to extract the data practice, and adjusted the similarity threshold to find the best extraction method. 

Figure \ref{fig:extraction} shows the best method is the overlap coefficient when the similarity threshold is 1 with an accuracy rate of 97.5$\%$. But before the 0.5 threshold, this effect is not as good as Euclidean distance and cosine distance using vector space for similarity calculation.

We think the reason is related to language expression habits. Words related to data processing tend to be specific and distinct. Such words do not undergo much change when they are used, usually only adding some modified words, which will not affect the original form of the words. For example, "\textit{city}" falls under the "\textit{address}", and the privacy policy would specify the scope with "\textit{currently living city}" or "\textit{working city}", these modifiers do not change the word "\textit{city}". This means that once the word "\textit{city}" is matched, we can assume that it describes an address-related data with a high probability. Therefore, overlap coefficient that is more inclined to include matching method perform best. However, with a low similarity threshold, this matching method introduces many erroneous results. For example, "\textit{mobile phone number}" and "\textit{ID number}" are grouped together because they both have the word "\textit{number}".

\noindent\fbox{\parbox[c]{0.95\linewidth}{
    \textit{\textbf{Answer to RQ3:} On 2136 calibrated sentences, the accuracy rate of sentence classification model is 97.98$\%$ , and the callback rate is 92.56$\%$. On 306 more finely calibrated sentences, the extraction method has an accuracy of 97.5$\%$. The method can effectively classify sentences and extract data-processing-related descriptions.}
}}

\subsection{Data Practice Distribution in Privacy Policy} 
\begin{figure}
    \setlength{\abovecaptionskip}{5pt}
    \setlength{\belowcaptionskip}{0pt}
    \begin{minipage}{0.65\linewidth}
        \flushleft
        \subfigure[Data Type Distribution]{
            \includegraphics[width=0.45\linewidth]{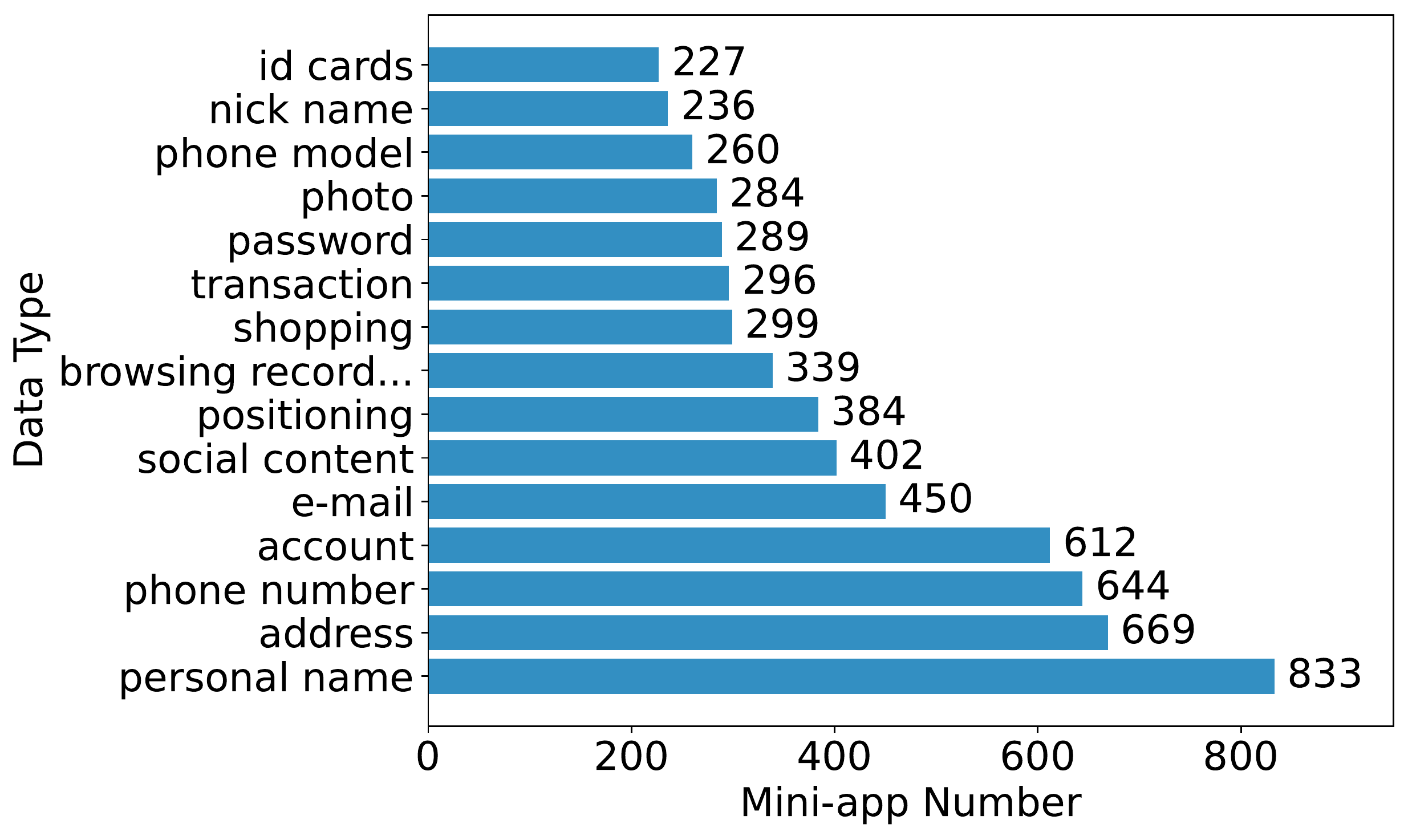}
            \label{fig:policydistribution-a}
        }
        \subfigure[Data Operation Distribution]{
            \includegraphics[width=0.45\linewidth]{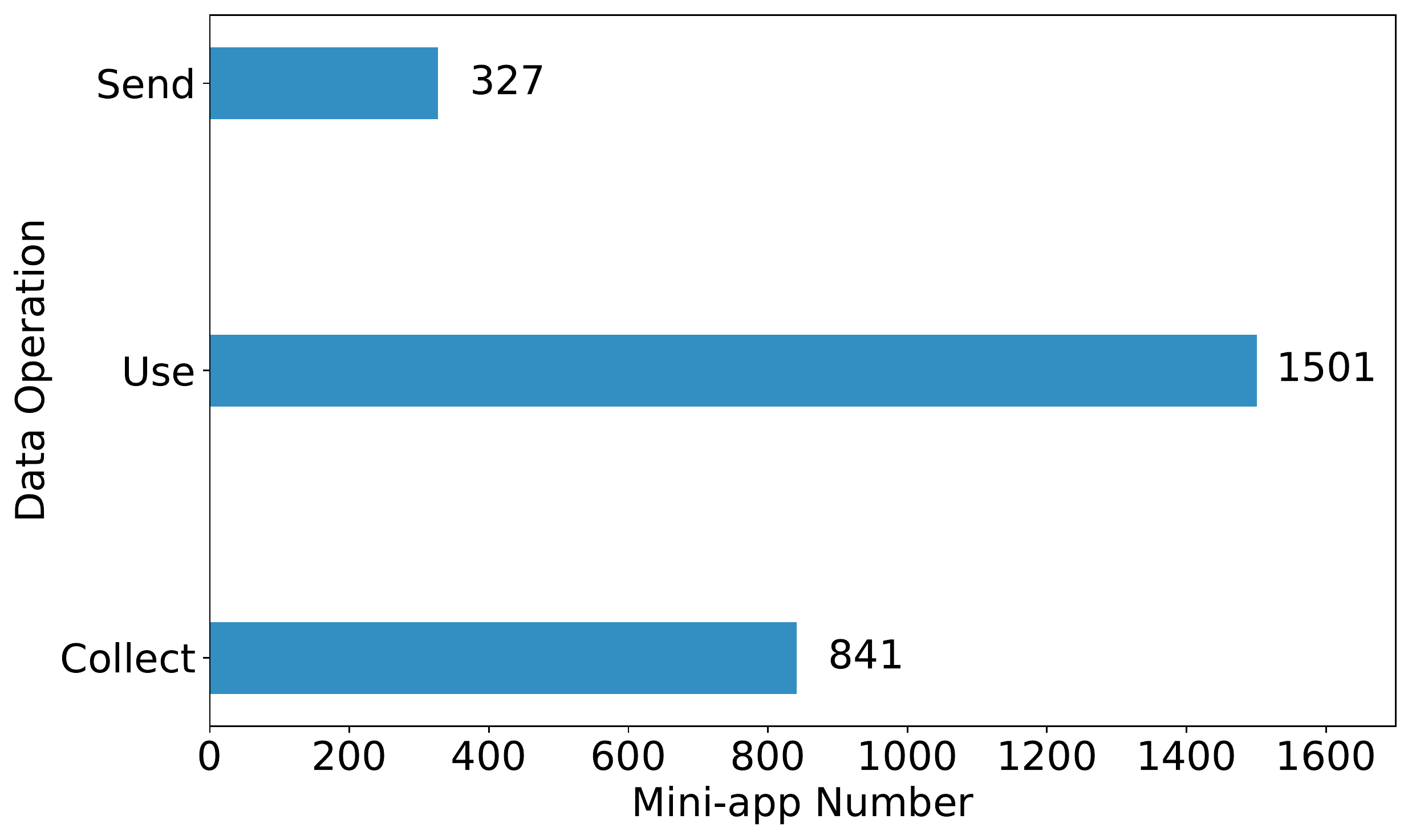}
            \label{fig:policydistribution-b}
        }
    \end{minipage}
    \begin{minipage}{0.16\linewidth}
        \subfigure[Data Practice Heatmap]{
            \includegraphics[width=\linewidth]{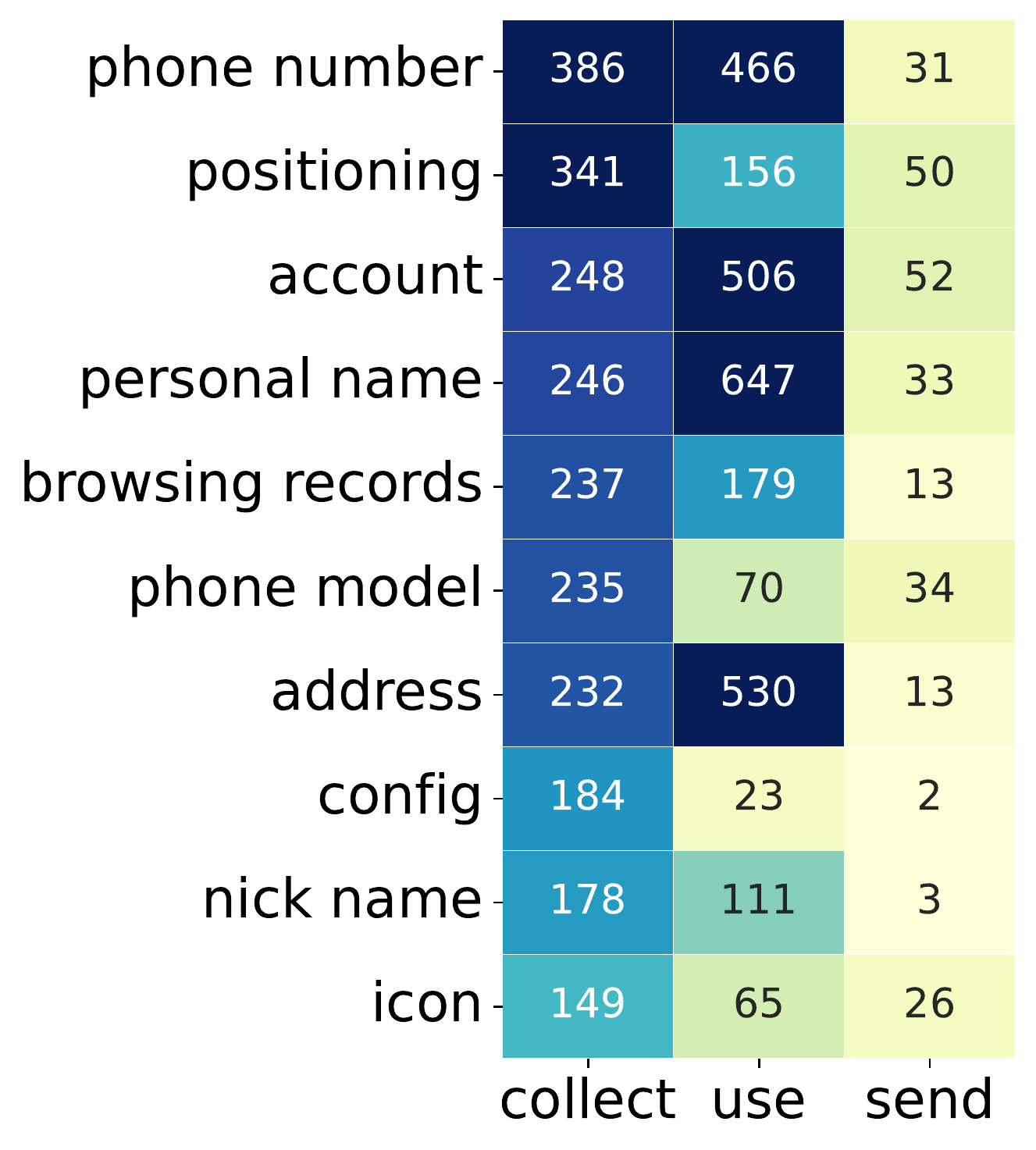}
            \label{fig:policydistribution-c}
        }
   \end{minipage}
   \caption{Data practice distribution in privacy policy.}
   \label{fig:policydistribution}
   \vspace{-20pt}
\end{figure}

We extracted data practices from 2,998 privacy policies. Among the results, 1,272 (42.43$\%$) did not contain relevant descriptions of data processing, and 1,746 (58.24$\%$) privacy policies included at least one data practice. Their distribution is shown in Figure \ref{fig:policydistribution}. Through statistics, we have the following findings:

(1) The data types stated in the privacy policy are wide-ranging, and most of them are user-sensitive information. As shown in Figure \ref{fig:policydistribution-a}, privacy policies favor data practices that state an individual's name, address, and mobile phone number. The difference from the code-side analysis results is that the privacy policy seldom states communication-related information, such as login credentials.

(2) The privacy policy states far less about collecting and sending than collecting. For the data operations shown in Figure \ref{fig:policydistribution-b}, there are only 841 privacy polices that mention sending data and 327 privacy polices that mention collecting data, accounting for 48.17$\%$ and 18.73$\%$. 

(3) The lack of privacy policy description is serious. On the one hand, only about half (58.24$\%$) of the privacy policies contain at least one data practice. On the other hand, as shown in Figure \ref{fig:policydistribution-c}, among the 1,746 privacy policies with data practices, the number of data practices  with the most declarations \textit{(personal name, Use)} is only 647.

\noindent\fbox{\parbox[c]{0.95\linewidth}{
    \textit{\textbf{Answer to RQ4:} Privacy policies tend to declare user sensitive information such as personal name (27.8$\%$), address (22.3$\%$), and phone number(21.5$\%$). They rarely declare the collection and sending of data. The percentages of collecting, using and sending statements are 28.1$\%$, 50.1$\%$ and 10.9$\%$. Half of the privacy policies do not mention data practices, and the number of most mentioned data practices \textit{(personal name, Use)} is 647.}
}}

\subsection{Inconsistency Distribution}

\begin{figure}
    \subfigure[Pattern Distribution]{
        \includegraphics[width=0.3\linewidth]{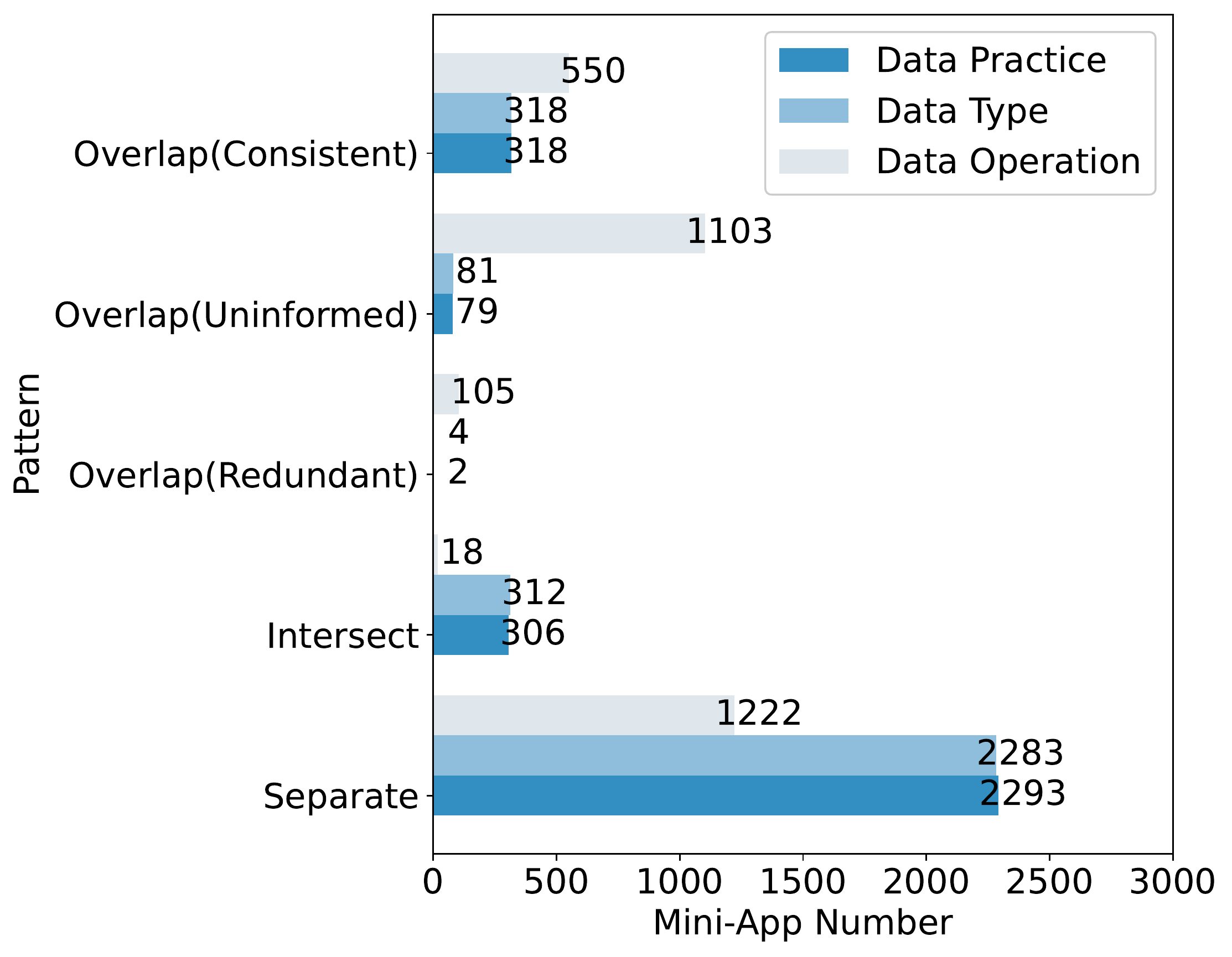}
        \label{fig:inconsistencydistribution-a}
    }
    \subfigure[Strong (Uninformed)]{
        \includegraphics[width=0.15\linewidth]{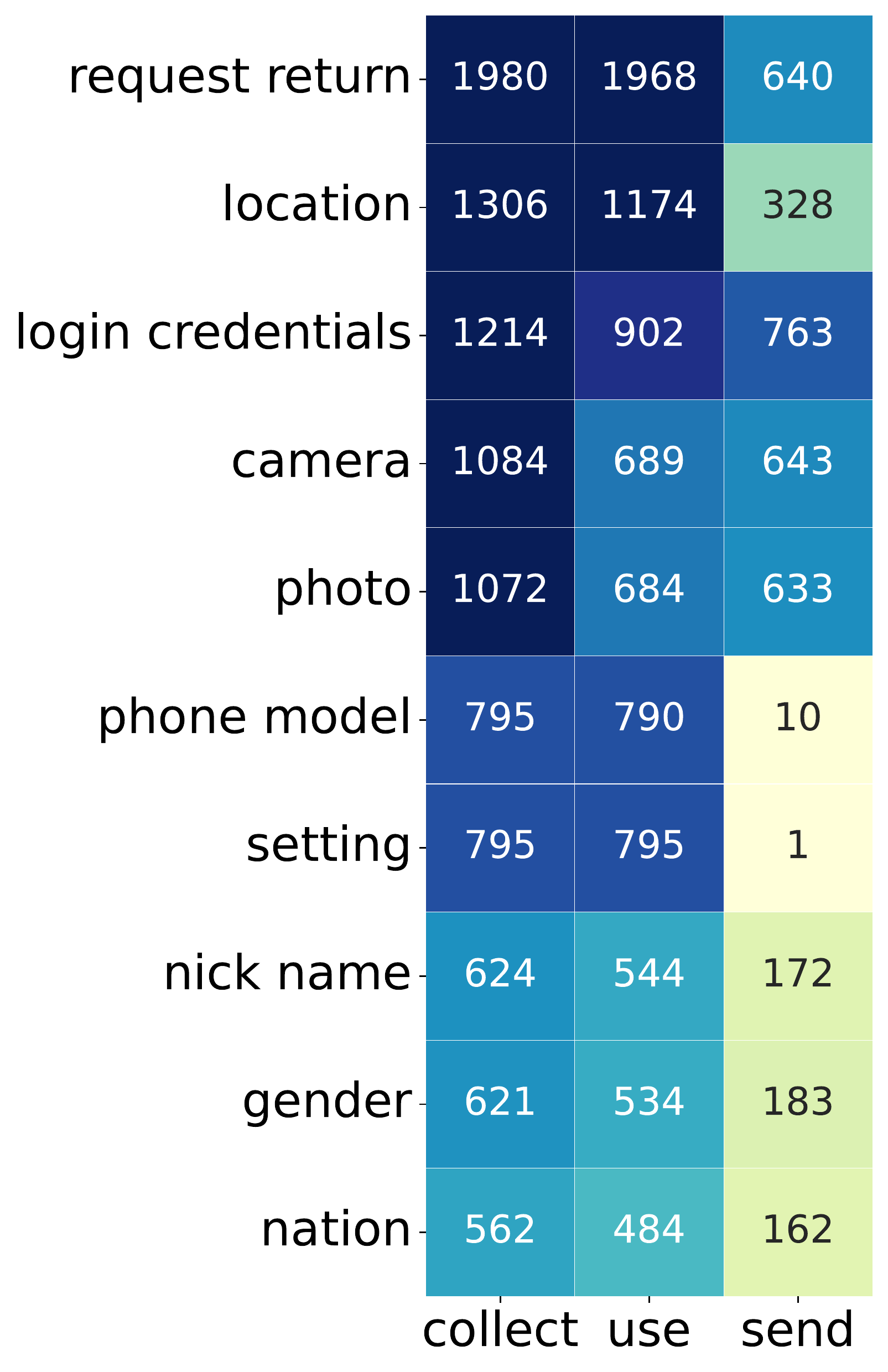}
        \label{fig:inconsistencydistribution-b}
    }
    \subfigure[Strong (Redundant)]{
        \includegraphics[width=0.15\linewidth]{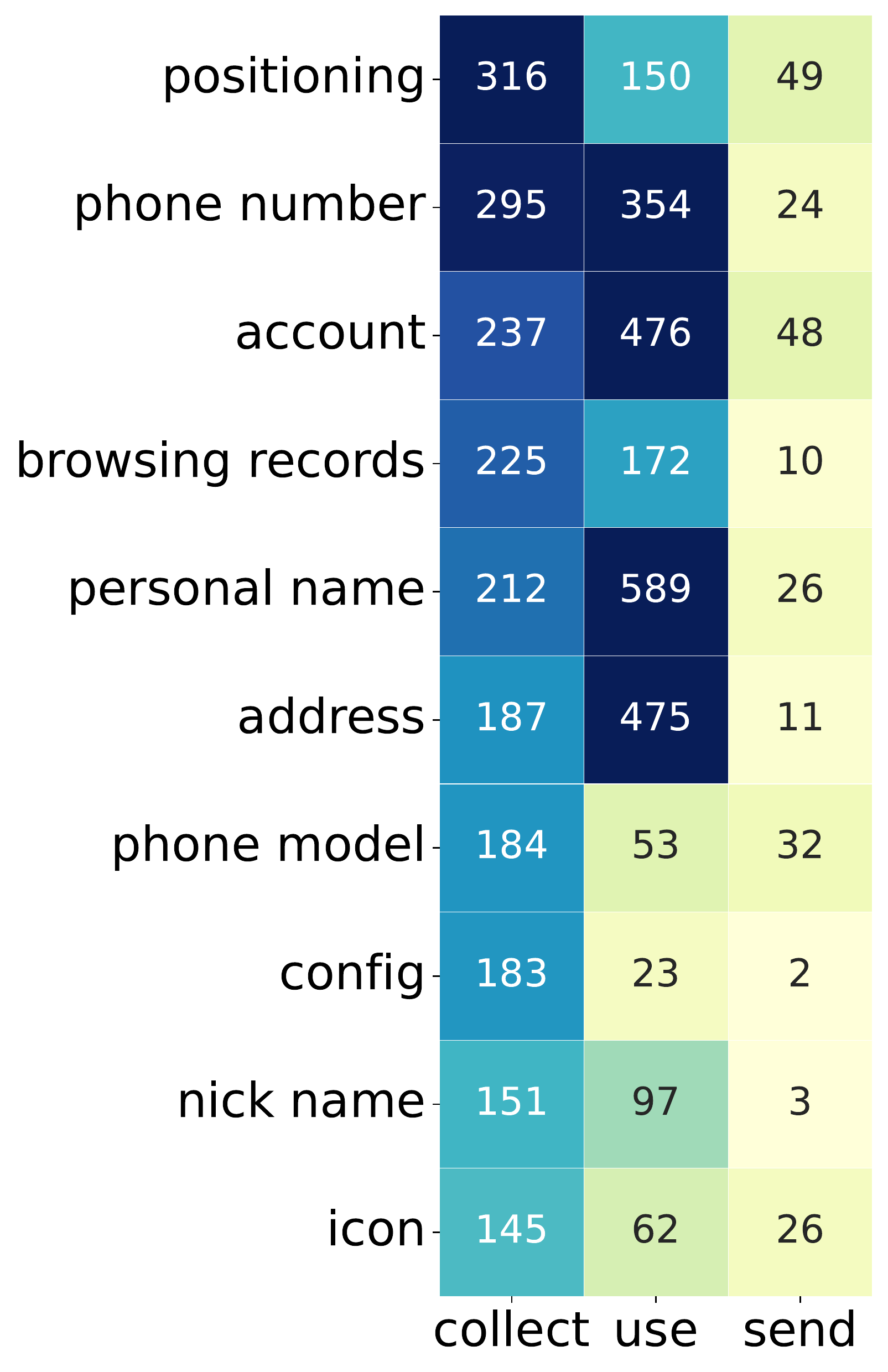}
        \label{fig:inconsistencydistribution-c}
    }
    \subfigure[Weak (Uninformed)]{
        \includegraphics[width=0.15\linewidth]{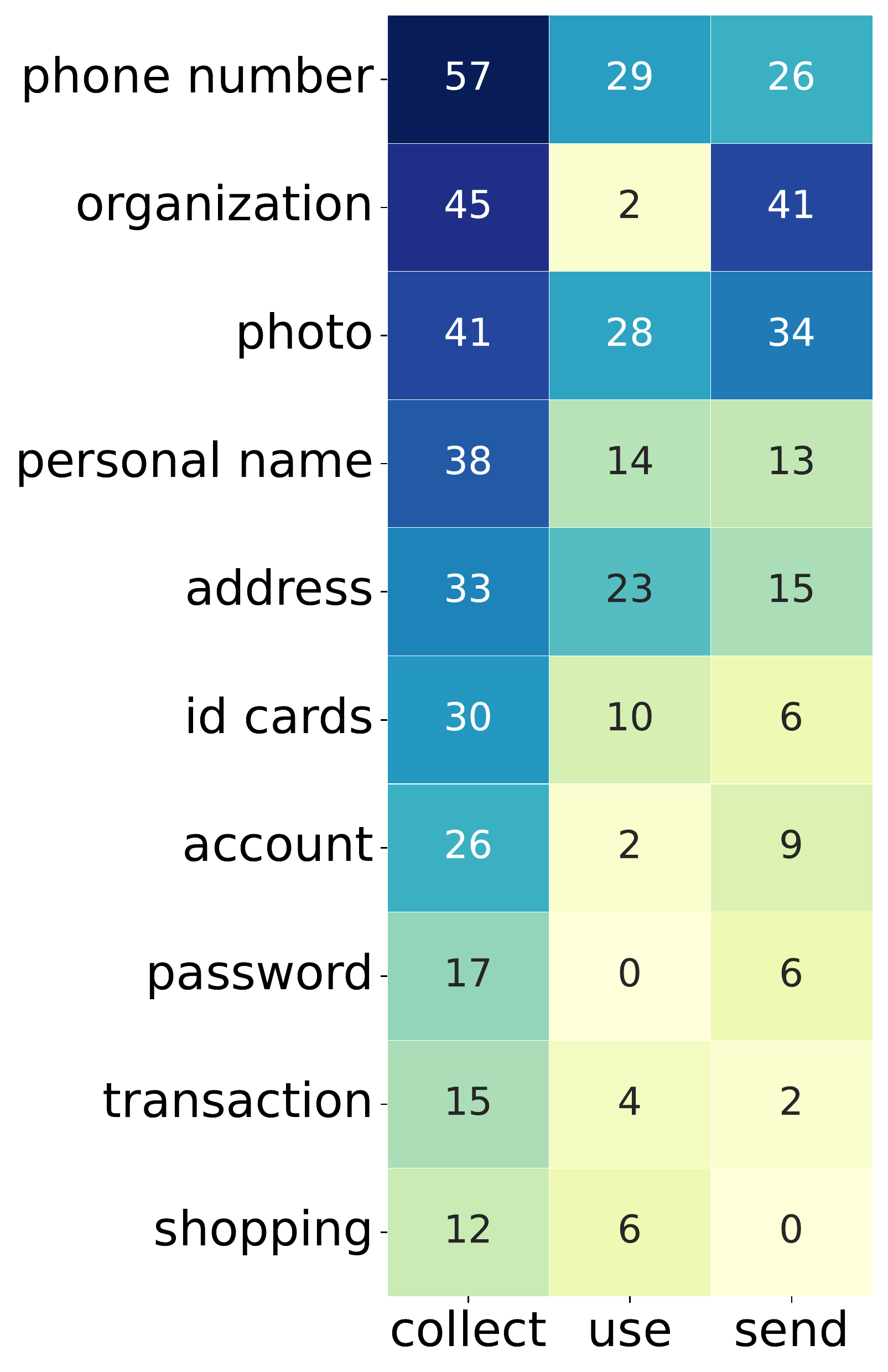}
        \label{fig:inconsistencydistribution-d}
    }
    \subfigure[Weak (Redundant)]{
        \includegraphics[width=0.15\linewidth]{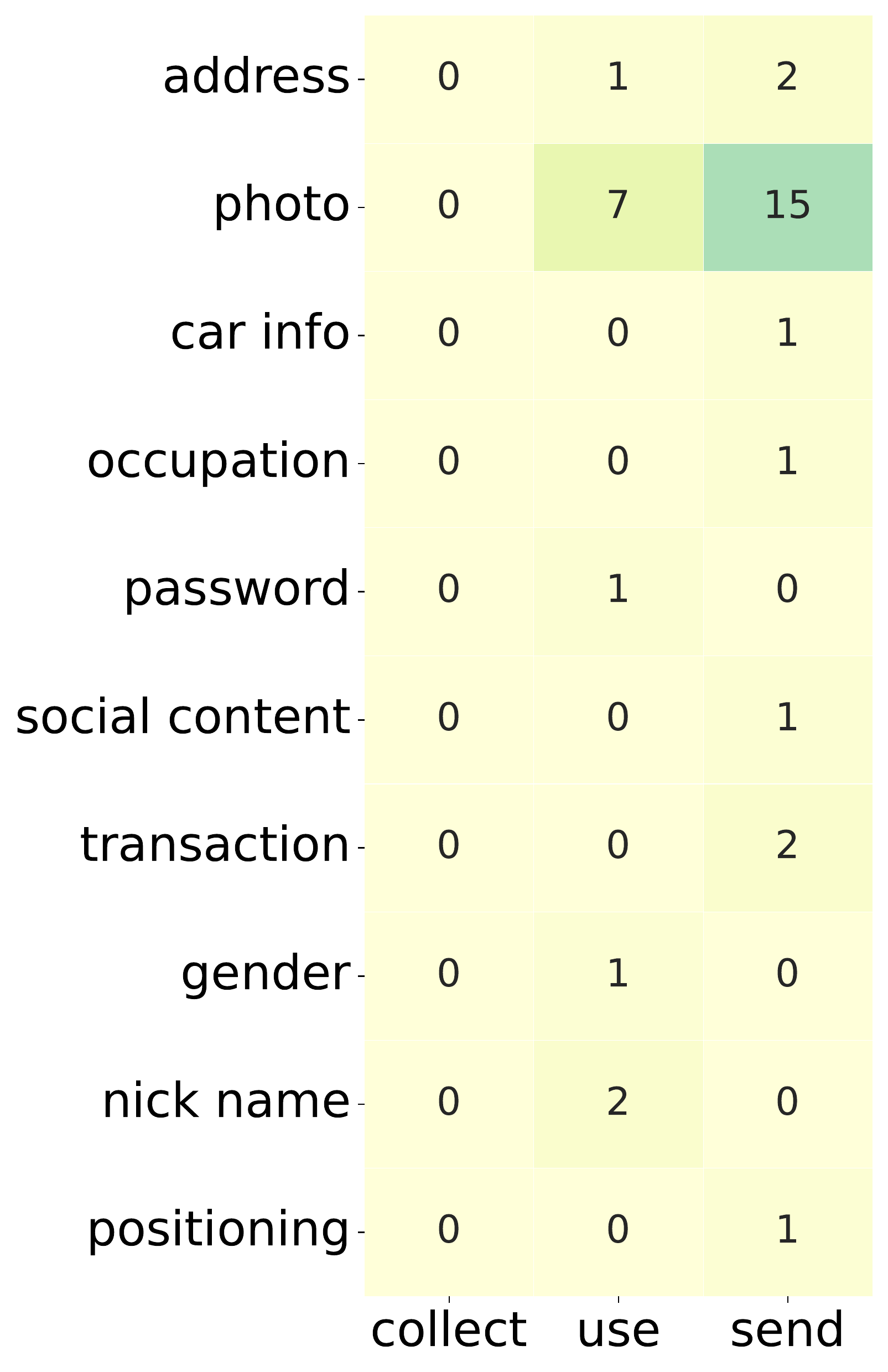}
        \label{fig:inconsistencydistribution-e}
    }
    \caption{Inconsistency distribution in mini-apps. (a) is the number distribution of the five patterns, and (b)-(e) are the top 10 data practices' heatmaps for strong and weak inconsistencies.}
    \label{fig:inconsistencydistribution}
    \vspace{-10pt}
\end{figure}

We performed consistency analysis on 2998 mini-apps and take three sets including data practices, data types and data operations to analyze the five patterns referred in Figure \ref{fig:consistencymode}. Figure \ref{fig:inconsistencydistribution-a} shows the distribution of five patterns. Figure \ref{fig:inconsistencydistribution-b}-\ref{fig:inconsistencydistribution-e} shows the top 10 data practices' heat maps for strong and weak inconsistencies. Through statistics, we have the following findings:

(1) The inconsistency in the mini-apps is very serious. Figure \ref{fig:inconsistencydistribution-a} shows that this inconsistency is mainly manifested in the separation of program code and privacy policy. The number of this most pattern is 2293, accounting for 76.5$\%$. For all four inconsistent patterns, the sum is as high as 2680, accounting for 89.4$\%$.

(2) These inconsistencies are mainly caused by inconsistent data types. Figure \ref{fig:inconsistencydistribution-a} shows that the pattern distributions of data types and data practices basically coincide. 

(3) These inconsistencies are mainly caused by the data practice of the program code exceeding the privacy policy. As shown in Figure \ref{fig:inconsistencydistribution-b}-\ref{fig:inconsistencydistribution-e}, inconsistencies are mainly distributed as strong uninformed inconsistencies. The largest uninformed data practice \textit{(request return, Collect)} appear in 1,980 mini-apps, and the largest redundant data practice \textit{(account, Use)} only appear in 589 mini-apps. The numbers of weakly inconsistent data practices are all less than 100.

\noindent\fbox{\parbox[c]{0.95\linewidth}{
    \textit{\textbf{Answer to RQ5} Among the 2998 mini-apps tested, there are 2,680 inconsistent mini-apps, accounting for 89.4$\%$. The inconsistency in mini-apps is extremely serious and is mainly caused by the data practice on the code side.}
}}

\section{Cases and Suggestions}
\label{sec:cases}

\subsection{Cases}

\begin{table}[]
    \centering
    \setlength{\abovecaptionskip}{5pt}
    \setlength{\belowcaptionskip}{0pt}
    \caption{Cases from the dataset.}
    \label{tab:cases}
    \resizebox{\textwidth}{!}{%
    \footnotesize
    \begin{tabular}{ccccc}
    \toprule
    \textbf{Basic Information} & \multicolumn{2}{c}{\textbf{Data practice*}} & \multicolumn{2}{c}{\textbf{Inconsistency}} \\
    \midrule
    \begin{tabular}[c]{@{}c@{}}ID\#Name\\ \#Downloads\#Category\end{tabular} & Program Code & Privacy Policy & \begin{tabular}[c]{@{}c@{}}Pattern\\ (Data Practice)\end{tabular} & Strength \\
    \midrule
    \begin{tabular}[c]{@{}c@{}}1\#Hui Charging\\ \#40704\#Service\end{tabular} & \begin{tabular}[c]{@{}c@{}}(camera, C\&U\&S),(request return, C\&U\&S),\\ (login credentials, C\&U\&S),(location, C\&U),\\ (photo, C\&S), (phone number, C\&U),\\ (sms, C\&U), (password, C\&U), \\ (browsing records, C\&U), (address, Ct\&U)\end{tabular} & - & Overlap(Uninformed) & Strong \\
    \midrule
    \begin{tabular}[c]{@{}c@{}}2\#KuaiShao\\ \#4598\#Tool\end{tabular} & (location,C\&U),(phone number,C\&U) & (account,U),(password,U),(express,U) & Seperation & Strong \\
    \midrule
    \begin{tabular}[c]{@{}c@{}}3\#Hangzhou Machinery \\ Application Platform\\ \#\textless{}100\#Government\end{tabular} & (location, C\&U) & (account,U), (password, U) & Seperation & Strong \\
    \midrule
    \begin{tabular}[c]{@{}c@{}}4\#Smart Canteen\\ \#\textless{}100\#Service\end{tabular} & \begin{tabular}[c]{@{}c@{}}(request return, C\&U), (address, C\&U),\\ (click record, C\&U)\end{tabular} & \begin{tabular}[c]{@{}c@{}}(account,U),(password,U),\\ (address, U), (e-mail, U)\end{tabular} & Intersection & Strong\&Weak \\
    \midrule
    \begin{tabular}[c]{@{}c@{}}5\#MeiYou\\ \#80743\#Service\end{tabular} & \begin{tabular}[c]{@{}c@{}}(request return,C\&U),(setting, C\&U), \\ (browing records, C\&U),(phone number, C\&U)\end{tabular} & \begin{tabular}[c]{@{}c@{}}(phone number,C\&U),(e-mail, C\&U),\\ (nick name, U), (icon, U), (app list, U),\\ (menstruation, C\&U), (browsing records, C\&U),\\ (personal name, U)\end{tabular} & Intersection & Strong \\
    \midrule
    6\#HuoShuiBao\#\textless{}100\#Service & - & - & Overlap(Consistent) & - \\
    \bottomrule
    \multicolumn{5}{l}{*Data practices with the same data type have been merged. For the data operations, "C" represents "Collect", "U" represents "Use" and "S" represents "Send".}\\
    \multicolumn{5}{l}{- means the result is empty.}
    \end{tabular}%
    }
    \vspace{-15pt}
    \end{table}

We randomly selected 6 real-world samples and their detailed results are shown in Table \ref{tab:cases}. Through sample analysis, we believe that the reasons for the inconsistencies are as follows:

\textbf{Missing description and rough category.} For case 1, its privacy policy simply summarizes user data as \textit{"registration data"}, and does not specifically describe how each data is acquired and used, which makes it impossible for the classifier to recognize it as a statement related to data processing. The privacy policy of case 2 mentions the collection of \textit{"express"} data, however, the scope of \textit{"express"} data is very broad and may include the location and mobile phone number. In fact, its privacy policy doesn't go any further to explain what part of the data is collected.

\textbf{Homogenized template and omitted characteristics.} Data practices in program code are hardly covered perfectly in the privacy policy. The complete mismatch in case 3 (the intersection between program code and privacy policy is empty) indicates that its privacy policy is absolutely separate from the actual mini-app behavior. Upon inspection, we find that its privacy policy uses a web-popular template, but omits the description of its own characteristics.

\textbf{Copied document and less adjustment.} Some mini-apps have many data practices descriptions in their privacy polices, and their redundancy is very high. For example, the privacy policy of case 5 mentions 12 data practices but there are only 8 data practices in its program code. Further, we found that its privacy policy is very similar to that of the Android application with the same name. However, this mini-app has only a fraction of the functions of an Android application with the same name, which causes the description to exceed the actual behavior.

In our opinions, these phenomena are closely related to the characteristics and ecology of mini-apps: 

(1) Mini-apps are small in size and low in development cost, which attracts many individual developers. They do not have professional knowledge of privacy policy writing, and the description documents are often written very briefly and irregularly.  This problem is also common in Android apps\cite{sunyaev2015policystatistic, story2019compliance, zimmeck2019maps}. Not enough experience results in the absence of a description of data processing in the privacy policy.

(2) There are also many mini-apps migrated from Android apps. They have partial functionality, but carry the full privacy policy on the Android side. The developers ignored the changes in the migration process, resulting in a lot of redundancy in the privacy policy.

\subsection{Suggestions}

Therefore, based on the above discussion, to improve such problems, we propose the following suggestions:

\textbf{Developer: Reduce the collection of irrelevant information and the straightforward use of templates.} Many mini-apps collect information that is not relevant to their functionality. For example, the function of case 3 is to issue a mechanical use application. There is no need to obtain the user's location information and its privacy policy also does not mention this. Therefore, reducing the collection of unnecessary information can reduce the risk of non-disclosure, and make the content of the privacy policy more concise. Case 6 is a mini-app that just displays text. It does not collect any data, and its privacy policy does not need to mention data processing-related content, so it meets the consistency requirement. In addition, it is recommended that developers learn the standard privacy policy requirements, reduce the direct use of templates, and provide tailor-made privacy policies for their own mini-apps.

\textbf{Platform: Provide data flow detection tools and privacy policy writing support.} For the platform, verifying whether the code contains data processing behaviors and feeding back the results to the developers can effectively help them reduce omissions in writing privacy policies\cite{balebako2014privacy}. Additionally, the submission of a privacy policy should be required. The platform can provide writing support and use the aforementioned analysis results to help developers improve their privacy policies.\vspace{-5pt}
\section{Threats to Validity}
\label{sec:threats}

\textbf{Tool independence.}
We use many open source tools to build such a consistency analysis system, but we do not conduct a detailed evaluation of the performance of these tools. If the results given by these tools are not complete, that may further affect the system's accuracy.

\textbf{Migration to other platforms.}
At present, many Android application platforms have the function of mini-apps. We briefly surveyed the popular mini-app platforms, including Alipay\cite{Alipay}, Baidu\cite{BaiduSmartProgram} and Bytedance\cite{BytedanceMiniApp}. They all use javascript as the core logic language, and a framework structure using layout files (xml-like format), style files (css-like format) and configuration files (json format). However, they use differentiated API lists and extended functionality, which can cause migration difficulties. For example, the APIs of the Alipay mini-app use \textit{my} as the main object name (WeChat mini-apps use \textit{wx}) and the Baidu provides many AI model calling interfaces in its API list.

\textbf{Migration to other regulation requirements.}
This consistency detection system mainly only verifies the user's right to know in laws and regulations. But there are also other compliance requirements, such as completeness\cite{tesfay2018privacyguide, rahman2022permpress}. It is necessary to classify and subclause privacy policies in more detail to detect whether they meet these requirements.

\vspace{-5pt}
\section{Related Work}
\label{sec:relatedwork}

\subsection{Mini-app Related Research}

Mini-apps are an emerging application form, and there are only a few related researches, which can be divided into vulnerabilities, testing and statistics. Lu et al.\cite{lu2020demystifying} reveal new security vulnerabilities in the new mini-app ecosystem by analyzing system resource exposure, sub-window deception, and sub-program hijacking. These vulnerabilities could allow an attacker to escalate privileges or access resources. At the same time, they evaluat the proposed attack on 11 popular mini-app platforms, revealing the prevalence of the security vulnerability. Liu et al.\cite{liu2020industry} design WeJalangi, a dynamic analysis tool for WeChat mini-apps, which can effectively mine mini-app code defects. MiniCrawler\cite{zhang2021measurement} injects and modifies the mini-app host platform application through the plug-in and filters and parses the field information returned by the download to obtain the link. They make measurements on the code ecology based on the crawled datasets, such as API usage and obfuscation rate. Wang et al.\cite{wang2022characterizing} summarized the main reasons for common bugs in WeChat mini-apps, and developed WeDetector to detect these bugs.

The above research does not focus on the analysis of the privacy in mini-apps. Also unfortunately, due to the security restrictions of the host app, some tools are no longer available (e.g. the interface used by MiniCrawler is invalid due to WeChat forced version updates, and WeJalangi is closed source).

\subsection{JavaScript Taint Analysis}

Unlike Android app, which uses Java, the mini-app adopts a web-like architecture and uses JavaScript as the core language. However, JavaScript's dynamic characteristics have brought many difficulties to language parsing. TAJS\cite{jensen2011modeling} is a classic work in JavaScript data flow analysis. It proposes an extending static analysis model for HTML DOM and browsers API, and applies it to Web applications. JSAI\cite{kashyap2014jsai} is a set of methods for converting JavaScript into an intermediate language (IR), providing definitions of precise and abstract semantics and unite them to give a configurable analysis. SAFE\cite{park2017analysis} is an parser that supports JavaScript abstract syntax tree rewriting and control flow graph analysis. JsPrime\cite{JsPrime} is a taint analysis tool for XSS attacks, uses the source and sink taint API lists in popular JavaScript libraries, and supports variable and function tracking. CodeQL\cite{CodeQL} is a code audit tool launched by Github. It first parses the code into a query database and then allows users to customize the query format. CodeQL also provides the taint analysis function for JavaScript. JsFlow\cite{hedin2014jsflow} performs information flow track and inspection through dynamic analysis. Understand\cite{Understand} is a code visualization tool which supports code dependency analysis and control flow graph analysis.

We compared the adaptability of the above analysis frameworks on mini-apps, as shown in Table \ref{javascript_tool}. The above work has been well adapted to some scenarios of JavaScript (e.g. Web for browsers and Node for servers), but it is still hard to achieve accurate analysis on mini-apps. As shown in the table, some tools provide limited interfaces and do not support taint analysis because these tools are designed for type checking rather than taint tracking. In addition, there are many asynchronous APIs and a new UI interaction mechanism which will affect the data flow analysis.

\begin{table}[t]
    \setlength{\abovecaptionskip}{5pt}
    \setlength{\belowcaptionskip}{0pt}
    \caption{Javascript analysis tools comparison}
    \centering
    \footnotesize
    \begin{tabular}{ccccccc}
    \toprule
    \multirow{2}{*}{Tool} & \multicolumn{3}{c}{Availabel Interfaces} & \multicolumn{3}{c}{Supportive} \\
     & Control Flow Analysis& Dependency Resolution & Data Flow Analysis & Web$^1$ & Node$^2$ & Mini-app \\ 
    \midrule
    TAJS & \checkmark & \checkmark & \checkmark & \checkmark & $\times$ & $\times$ \\
    JSAI & $\times$ & $\times$ & $\times$ & \checkmark & $\times$ & $\times$ \\
    SAFE & \checkmark & $\times$ & $\times$ & \checkmark & \checkmark & $\times$ \\
    JsPrime & $\times$ & \checkmark & \checkmark & \checkmark & \checkmark & $\times$ \\
    CodeQL-Js & \checkmark & \checkmark & \checkmark & \checkmark & \checkmark & $\times$ \\
    JsFlow & $\times$ & $\times$ & \checkmark & \checkmark & $\times$ & $\times$ \\
    Understand-Js & \checkmark & \checkmark & $\times$ & \checkmark & \checkmark & $\times$ \\ \bottomrule
    \multicolumn{7}{l}{\footnotesize 1. The web application processes HTML files, based on the DOM interface, and runs in a browser environment.}\\
    \multicolumn{7}{l}{\footnotesize 2. The node application is based on Chrome JavaScript, event-driven, and runs on the server.}\\
    \end{tabular}%
    \label{javascript_tool}
    \vspace{-15pt}
    \end{table}

\subsection{Compliance Analysis on Android Apps}

Mobile applications' privacy and security issues are a common topic, and many laws and regulations promulgated have placed restrictions on data processing. However, the behavior of the software is usually imperceptible to the user, and the correctness of the documents provided by the software, like privacy policies, can hardly be verified\cite{yu2016can}. The consistency problem is to use program analysis methods to obtain how the program collects and uses user data information, and compares this result with the description text to check whether there are any violations or omissions.

\textbf{Behavior analysis.} According to different analysis objects, behavior analysis can be divided into analysis based on tainted data flow and analysis based on user interface. The analysis based on tainted data flow obtains the data dependencies in the program code through static or dynamic methods, and discovers the behavior of the program when processing data. The target of this type of work is to study accurate data flow search methods under a specific code environment or architecture. The related work includes Flowdroid\cite{arzt2014flowdroid}, IccTA\cite{li2015iccta},TaintDroid\cite{enck2014taintdroid}, DroidSafe\cite{gordon2015information} and Amandroid\cite{wei2018amandroid}. The analysis based on the user interface provides additional explanatory support for data breaches by extracting textual information on the UI, such as determining the type of user input data. The related work includes UIPicker\cite{nan2015uipicker}, Uiref\cite{andow2017uiref}, Bidtext\cite{huang2016detecting} and IconIntent\cite{xiao2019iconintent}.

These studies provide ideological guidance for the behavior analysis of mini-apps. However, they cannot provide sufficient tool support because mini-apps and apps employ two completely different programming languages (Javascript and Java). In addition, in the process of layout resource resolution, the Web-like structure in mini-app is also different from that of Android.

\textbf{Description analysis.} CHABADA\cite{gorla2014chabada} is an early research to detect the difference between APP behavior and description. It clusters the keywords that appear in different types of applications, counts the frequency of different APIs, and uses a classifier to detect anomalies. Autocog\cite{qu2014autocog} expands the permission set of existing research and uses learning methods to improve the correlation accuracy of behavior matching. AsDroid\cite{huang2014asdroid} uses static program analysis and interface text to detect conflicts. Polisis\cite{harkous2018polisis} proposes an automated framework based on deep learning to achieve query and question answering of privacy description documents. PPChecker\cite{yu2018ppchecker} proposes a privacy policy evaluation method that evaluates privacy policies around completeness, correctness, consistency, and user-friendliness. PoliCheck\cite{andow2020policheck} combines AppCensus\cite{AppCensus} to obtain data streams, uses PolicyLint\cite{andow2019policylint} to parse privacy policy documents, and finally builds a flow-to-policy model to detect consistency. Purliance\cite{bui2021purliance} complements the purpose analysis of the data usage section, using network traffic to represent the specific behavior of the purpose. 

Some of the above studies have coarse granularity in the extraction of program behavior, and only extract simple program behavior elements such as API usage frequency in CHABADA\cite{gorla2014chabada} or traffic in Purliance\cite{bui2021purliance}. In addition, the scope of data types and data operations for some comparison work is limited. For example, PolicyCheck\cite{andow2020policheck} only analyzed 12 data types supported by the data flow analysis tools AppCensus\cite{AppCensus}.

\vspace{-5pt}
\section{Conclusion}
\label{sec:conclusion}

This paper proposes a mini-apps consistency analysis framework for the first time. The system contains four procedures: preparation, identification of data practice in program code and privacy policy, and consistency comparison. We crawl 100,000 mini-apps on WeChat client in the wild and extract 2,998 with a privacy policy. Among them, only 318 meet the consistency requirements, 2,680 are inconsistent, and the proportion of inconsistencies is as high as 89.4$\%$. Based on the above results and 6 anlyzed real-world cases, we think the reasons for the inconsistencies mainly caused by mini-apps' characteristics and ecology. Developers are suggested to reduce the collection of irrelevant information and the straightforward use of templates, and the platform is suggested to provide data flow detection tools and privacy policy writing support.

\bibliographystyle{ACM-Reference-Format}
\bibliography{reference.bib}


\begin{thebibliography}{67}


\ifx \showCODEN    \undefined \def \showCODEN     #1{\unskip}     \fi
\ifx \showDOI      \undefined \def \showDOI       #1{#1}\fi
\ifx \showISBNx    \undefined \def \showISBNx     #1{\unskip}     \fi
\ifx \showISBNxiii \undefined \def \showISBNxiii  #1{\unskip}     \fi
\ifx \showISSN     \undefined \def \showISSN      #1{\unskip}     \fi
\ifx \showLCCN     \undefined \def \showLCCN      #1{\unskip}     \fi
\ifx \shownote     \undefined \def \shownote      #1{#1}          \fi
\ifx \showarticletitle \undefined \def \showarticletitle #1{#1}   \fi
\ifx \showURL      \undefined \def \showURL       {\relax}        \fi
\providecommand\bibfield[2]{#2}
\providecommand\bibinfo[2]{#2}
\providecommand\natexlab[1]{#1}
\providecommand\showeprint[2][]{arXiv:#2}

\bibitem[GDP(2018)]%
        {GDPR}
 \bibinfo{year}{2018}\natexlab{}.
\newblock \bibinfo{title}{{General Data Protection Regulation}}.
\newblock \bibinfo{howpublished}{\url{https://gdpr-info.eu/}}.
\newblock
\newblock
\shownote{[Online; accessed 25-May-2018]}.


\bibitem[chi(2020)]%
        {chinesestandard}
 \bibinfo{year}{2020}\natexlab{}.
\newblock \bibinfo{title}{{Information security technology—Personal
  information security specification}}.
\newblock
  \bibinfo{howpublished}{\url{https://openstd.samr.gov.cn/bzgk/gb/newGbInfo?hcno=4568F276E0F8346EB0FBA097AA0CE05E}}.
\newblock
\newblock
\shownote{[Online; accessed 1-October-2020]}.


\bibitem[nan(2020)]%
        {nandureport}
 \bibinfo{year}{2020}\natexlab{}.
\newblock \bibinfo{title}{{MiniApp Personal Information Protection Research
  Report}}.
\newblock
  \bibinfo{howpublished}{\url{http://www.caict.ac.cn/kxyj/qwfb/ztbg/202006/P020200611352964581409.pdf}}.
\newblock
\newblock
\shownote{[Online; accessed 27-June-2020]}.


\bibitem[ald(2022)]%
        {aldarticle}
 \bibinfo{year}{2022}\natexlab{}.
\newblock \bibinfo{title}{{2021 MiniApp Internet Development White Paper}}.
\newblock
  \bibinfo{howpublished}{\url{https://aldzs.com/viewpointarticle?id=16175}}.
\newblock
\newblock
\shownote{[Online; accessed 1-January-2022]}.


\bibitem[Ali(2022)]%
        {Alipay}
 \bibinfo{year}{2022}\natexlab{}.
\newblock \bibinfo{title}{Alipay MiniApp Documentation Guidelines}.
\newblock \bibinfo{howpublished}{\url{https://opendocs.alipay.com/mini}}.
\newblock
\newblock
\shownote{[Online; accessed 1-January-2022]}.


\bibitem[App(2022)]%
        {AppCensus}
 \bibinfo{year}{2022}\natexlab{}.
\newblock \bibinfo{title}{{Are your mobile apps protecting user data? Automated
  app analysis helps you understand and verify your app's privacy behaviors.}}
\newblock \bibinfo{howpublished}{\url{https://www.appcensus.io/}}.
\newblock
\newblock
\shownote{[Online; accessed 1-January-2022]}.


\bibitem[Byt(2022)]%
        {BytedanceMiniApp}
 \bibinfo{year}{2022}\natexlab{}.
\newblock \bibinfo{title}{Bytedance MiniApp overview}.
\newblock
  \bibinfo{howpublished}{\url{https://developer.open-douyin.com/docs/resource/zh-CN/mini-app/introduction/overview}}.
\newblock
\newblock
\shownote{[Online; accessed 1-January-2022]}.


\bibitem[CCP(2022)]%
        {CCPA}
 \bibinfo{year}{2022}\natexlab{}.
\newblock \bibinfo{title}{California Consumer Privacy Act}.
\newblock \bibinfo{howpublished}{\url{https://oag.ca.gov/privacy/ccpa}}.
\newblock
\newblock
\shownote{[Online; accessed 1-January-2022]}.


\bibitem[Air(2022)]%
        {Airtest}
 \bibinfo{year}{2022}\natexlab{}.
\newblock \bibinfo{title}{Cross-Platform UI Automation Framework for Games and
  Apps}.
\newblock
  \bibinfo{howpublished}{\url{https://github.com/AirtestProject/Airtest}}.
\newblock
\newblock
\shownote{[Online; accessed 1-January-2022]}.


\bibitem[Cod(2022)]%
        {CodeQL}
 \bibinfo{year}{2022}\natexlab{}.
\newblock \bibinfo{title}{{Discover vulnerabilities across a codebase with
  CodeQL}}.
\newblock \bibinfo{howpublished}{\url{https://codeql.github.com/}}.
\newblock
\newblock
\shownote{[Online; accessed 1-January-2022]}.


\bibitem[MCD(2022)]%
        {MCDS}
 \bibinfo{year}{2022}\natexlab{}.
\newblock \bibinfo{title}{Miniapp Consistency Detection System}.
\newblock
  \bibinfo{howpublished}{\url{https://github.com/xjtu-intsoft/miniapp-consistency-detection}}.
\newblock
\newblock
\shownote{[Online; accessed 1-January-2022]}.


\bibitem[min(2022)]%
        {miniappwhitepaper}
 \bibinfo{year}{2022}\natexlab{}.
\newblock \bibinfo{title}{{MiniApp Standardization White Paper}}.
\newblock
  \bibinfo{howpublished}{\url{https://www.w3.org/TR/mini-app-white-paper/}}.
\newblock
\newblock
\shownote{[Online; accessed 1-July-2022]}.


\bibitem[sta(2022)]%
        {statisticsappnumber}
 \bibinfo{year}{2022}\natexlab{}.
\newblock \bibinfo{title}{{Number of available applications in the Google Play
  Store from December 2009 to March 2022}}.
\newblock
  \bibinfo{howpublished}{\url{https://www.statista.com/statistics/266210/number-of-available-applications-in-the-google-play-store/}}.
\newblock
\newblock
\shownote{[Online; accessed 27-July-2022]}.


\bibitem[Chi(2022)]%
        {ChinaProtectionLaw}
 \bibinfo{year}{2022}\natexlab{}.
\newblock \bibinfo{title}{Personal Information Protection Law of the People's
  Republic of China}.
\newblock
  \bibinfo{howpublished}{\url{http://www.npc.gov.cn/npc/c30834/202108/a8c4e3672c74491a80b53a172bb753fe.shtml}}.
\newblock
\newblock
\shownote{[Online; accessed 1-January-2022]}.


\bibitem[Bai(2022)]%
        {BaiduSmartProgram}
 \bibinfo{year}{2022}\natexlab{}.
\newblock \bibinfo{title}{Smart MiniApp Documentation}.
\newblock \bibinfo{howpublished}{\url{https://smartprogram.baidu.com/docs/}}.
\newblock
\newblock
\shownote{[Online; accessed 1-January-2022]}.


\bibitem[Und(2022)]%
        {Understand}
 \bibinfo{year}{2022}\natexlab{}.
\newblock \bibinfo{title}{{Understand By SciTools}}.
\newblock \bibinfo{howpublished}{\url{https://www.scitools.com/}}.
\newblock
\newblock
\shownote{[Online; accessed 1-January-2022]}.


\bibitem[wxa(2022)]%
        {wxappUnpacker}
 \bibinfo{year}{2022}\natexlab{}.
\newblock \bibinfo{title}{{WeChat MiniApp Unpacking Tool: wxappUnpacker}}.
\newblock
  \bibinfo{howpublished}{\url{https://github.com/xuedingmiaojun/wxappUnpacker}}.
\newblock
\newblock
\shownote{[Online; accessed 1-January-2022]}.


\bibitem[Andow et~al\mbox{.}(2017)]%
        {andow2017uiref}
\bibfield{author}{\bibinfo{person}{Benjamin Andow}, \bibinfo{person}{Akhil
  Acharya}, \bibinfo{person}{Dengfeng Li}, \bibinfo{person}{William Enck},
  \bibinfo{person}{Kapil Singh}, {and} \bibinfo{person}{Tao Xie}.}
  \bibinfo{year}{2017}\natexlab{}.
\newblock \showarticletitle{Uiref: analysis of sensitive user inputs in android
  applications}. In \bibinfo{booktitle}{\emph{Proceedings of the 10th ACM
  Conference on Security and Privacy in Wireless and Mobile Networks}}.
  \bibinfo{pages}{23--34}.
\newblock


\bibitem[Andow et~al\mbox{.}(2019)]%
        {andow2019policylint}
\bibfield{author}{\bibinfo{person}{Benjamin Andow},
  \bibinfo{person}{Samin~Yaseer Mahmud}, \bibinfo{person}{Wenyu Wang},
  \bibinfo{person}{Justin Whitaker}, \bibinfo{person}{William Enck},
  \bibinfo{person}{Bradley Reaves}, \bibinfo{person}{Kapil Singh}, {and}
  \bibinfo{person}{Tao Xie}.} \bibinfo{year}{2019}\natexlab{}.
\newblock \showarticletitle{$\{$PolicyLint$\}$: Investigating Internal Privacy
  Policy Contradictions on Google Play}. In \bibinfo{booktitle}{\emph{28th
  USENIX security symposium (USENIX security 19)}}. \bibinfo{pages}{585--602}.
\newblock


\bibitem[Andow et~al\mbox{.}(2020)]%
        {andow2020policheck}
\bibfield{author}{\bibinfo{person}{Benjamin Andow},
  \bibinfo{person}{Samin~Yaseer Mahmud}, \bibinfo{person}{Justin Whitaker},
  \bibinfo{person}{William Enck}, \bibinfo{person}{Bradley Reaves},
  \bibinfo{person}{Kapil Singh}, {and} \bibinfo{person}{Serge Egelman}.}
  \bibinfo{year}{2020}\natexlab{}.
\newblock \showarticletitle{Actions Speak Louder than
  Words:$\{$Entity-Sensitive$\}$ Privacy Policy and Data Flow Analysis with
  $\{$PoliCheck$\}$}. In \bibinfo{booktitle}{\emph{29th USENIX Security
  Symposium (USENIX Security 20)}}. \bibinfo{pages}{985--1002}.
\newblock


\bibitem[Arzt et~al\mbox{.}(2014)]%
        {arzt2014flowdroid}
\bibfield{author}{\bibinfo{person}{Steven Arzt}, \bibinfo{person}{Siegfried
  Rasthofer}, \bibinfo{person}{Christian Fritz}, \bibinfo{person}{Eric Bodden},
  \bibinfo{person}{Alexandre Bartel}, \bibinfo{person}{Jacques Klein},
  \bibinfo{person}{Yves Le~Traon}, \bibinfo{person}{Damien Octeau}, {and}
  \bibinfo{person}{Patrick McDaniel}.} \bibinfo{year}{2014}\natexlab{}.
\newblock \showarticletitle{Flowdroid: Precise context, flow, field,
  object-sensitive and lifecycle-aware taint analysis for android apps}.
\newblock \bibinfo{journal}{\emph{Acm Sigplan Notices}} \bibinfo{volume}{49},
  \bibinfo{number}{6} (\bibinfo{year}{2014}), \bibinfo{pages}{259--269}.
\newblock


\bibitem[Balebako et~al\mbox{.}(2014)]%
        {balebako2014privacy}
\bibfield{author}{\bibinfo{person}{Rebecca Balebako}, \bibinfo{person}{Abigail
  Marsh}, \bibinfo{person}{Jialiu Lin}, \bibinfo{person}{Jason~I Hong}, {and}
  \bibinfo{person}{Lorrie~Faith Cranor}.} \bibinfo{year}{2014}\natexlab{}.
\newblock \showarticletitle{The privacy and security behaviors of smartphone
  app developers}.
\newblock  (\bibinfo{year}{2014}).
\newblock


\bibitem[Bui et~al\mbox{.}(2021)]%
        {bui2021purliance}
\bibfield{author}{\bibinfo{person}{Duc Bui}, \bibinfo{person}{Yuan Yao},
  \bibinfo{person}{Kang~G Shin}, \bibinfo{person}{Jong-Min Choi}, {and}
  \bibinfo{person}{Junbum Shin}.} \bibinfo{year}{2021}\natexlab{}.
\newblock \showarticletitle{Consistency Analysis of Data-Usage Purposes in
  Mobile Apps}. In \bibinfo{booktitle}{\emph{Proceedings of the 2021 ACM SIGSAC
  Conference on Computer and Communications Security}}.
  \bibinfo{pages}{2824--2843}.
\newblock


\bibitem[Che et~al\mbox{.}(2020)]%
        {che2020n}
\bibfield{author}{\bibinfo{person}{Wanxiang Che}, \bibinfo{person}{Yunlong
  Feng}, \bibinfo{person}{Libo Qin}, {and} \bibinfo{person}{Ting Liu}.}
  \bibinfo{year}{2020}\natexlab{}.
\newblock \showarticletitle{N-LTP: A Open-source Neural Chinese Language
  Technology Platform with Pretrained Models}.
\newblock \bibinfo{journal}{\emph{arXiv preprint arXiv:2009.11616}}
  (\bibinfo{year}{2020}).
\newblock


\bibitem[Chugh et~al\mbox{.}(2009)]%
        {chugh2009staged}
\bibfield{author}{\bibinfo{person}{Ravi Chugh}, \bibinfo{person}{Jeffrey~A
  Meister}, \bibinfo{person}{Ranjit Jhala}, {and} \bibinfo{person}{Sorin
  Lerner}.} \bibinfo{year}{2009}\natexlab{}.
\newblock \showarticletitle{Staged information flow for JavaScript}. In
  \bibinfo{booktitle}{\emph{Proceedings of the 30th ACM SIGPLAN conference on
  programming language design and implementation}}. \bibinfo{pages}{50--62}.
\newblock


\bibitem[Enck et~al\mbox{.}(2014)]%
        {enck2014taintdroid}
\bibfield{author}{\bibinfo{person}{William Enck}, \bibinfo{person}{Peter
  Gilbert}, \bibinfo{person}{Seungyeop Han}, \bibinfo{person}{Vasant
  Tendulkar}, \bibinfo{person}{Byung-Gon Chun}, \bibinfo{person}{Landon~P Cox},
  \bibinfo{person}{Jaeyeon Jung}, \bibinfo{person}{Patrick McDaniel}, {and}
  \bibinfo{person}{Anmol~N Sheth}.} \bibinfo{year}{2014}\natexlab{}.
\newblock \showarticletitle{Taintdroid: an information-flow tracking system for
  realtime privacy monitoring on smartphones}.
\newblock \bibinfo{journal}{\emph{ACM Transactions on Computer Systems (TOCS)}}
  \bibinfo{volume}{32}, \bibinfo{number}{2} (\bibinfo{year}{2014}),
  \bibinfo{pages}{1--29}.
\newblock


\bibitem[Fan et~al\mbox{.}(2020)]%
        {fan2020empirical}
\bibfield{author}{\bibinfo{person}{Ming Fan}, \bibinfo{person}{Le Yu},
  \bibinfo{person}{Sen Chen}, \bibinfo{person}{Hao Zhou},
  \bibinfo{person}{Xiapu Luo}, \bibinfo{person}{Shuyue Li},
  \bibinfo{person}{Yang Liu}, \bibinfo{person}{Jun Liu}, {and}
  \bibinfo{person}{Ting Liu}.} \bibinfo{year}{2020}\natexlab{}.
\newblock \showarticletitle{An empirical evaluation of GDPR compliance
  violations in Android mHealth apps}. In \bibinfo{booktitle}{\emph{2020 IEEE
  31st International Symposium on Software Reliability Engineering (ISSRE)}}.
  IEEE, \bibinfo{pages}{253--264}.
\newblock


\bibitem[Gordon et~al\mbox{.}(2015)]%
        {gordon2015information}
\bibfield{author}{\bibinfo{person}{Michael~I Gordon}, \bibinfo{person}{Deokhwan
  Kim}, \bibinfo{person}{Jeff~H Perkins}, \bibinfo{person}{Limei Gilham},
  \bibinfo{person}{Nguyen Nguyen}, {and} \bibinfo{person}{Martin~C Rinard}.}
  \bibinfo{year}{2015}\natexlab{}.
\newblock \showarticletitle{Information flow analysis of android applications
  in droidsafe.}. In \bibinfo{booktitle}{\emph{NDSS}},
  Vol.~\bibinfo{volume}{15}. \bibinfo{pages}{110}.
\newblock


\bibitem[Gorla et~al\mbox{.}(2014)]%
        {gorla2014chabada}
\bibfield{author}{\bibinfo{person}{Alessandra Gorla}, \bibinfo{person}{Ilaria
  Tavecchia}, \bibinfo{person}{Florian Gross}, {and} \bibinfo{person}{Andreas
  Zeller}.} \bibinfo{year}{2014}\natexlab{}.
\newblock \showarticletitle{Checking app behavior against app descriptions}. In
  \bibinfo{booktitle}{\emph{Proceedings of the 36th international conference on
  software engineering}}. \bibinfo{pages}{1025--1035}.
\newblock


\bibitem[Harkous et~al\mbox{.}(2018)]%
        {harkous2018polisis}
\bibfield{author}{\bibinfo{person}{Hamza Harkous}, \bibinfo{person}{Kassem
  Fawaz}, \bibinfo{person}{R{\'e}mi Lebret}, \bibinfo{person}{Florian Schaub},
  \bibinfo{person}{Kang~G Shin}, {and} \bibinfo{person}{Karl Aberer}.}
  \bibinfo{year}{2018}\natexlab{}.
\newblock \showarticletitle{Polisis: Automated analysis and presentation of
  privacy policies using deep learning}. In \bibinfo{booktitle}{\emph{27th
  USENIX Security Symposium (USENIX Security 18)}}. \bibinfo{pages}{531--548}.
\newblock


\bibitem[He et~al\mbox{.}(2020)]%
        {he2020textexerciser}
\bibfield{author}{\bibinfo{person}{Yuyu He}, \bibinfo{person}{Lei Zhang},
  \bibinfo{person}{Zhemin Yang}, \bibinfo{person}{Yinzhi Cao},
  \bibinfo{person}{Keke Lian}, \bibinfo{person}{Shuai Li}, \bibinfo{person}{Wei
  Yang}, \bibinfo{person}{Zhibo Zhang}, \bibinfo{person}{Min Yang},
  \bibinfo{person}{Yuan Zhang}, {et~al\mbox{.}}}
  \bibinfo{year}{2020}\natexlab{}.
\newblock \showarticletitle{TextExerciser: feedback-driven text input
  exercising for android applications}. In \bibinfo{booktitle}{\emph{2020 IEEE
  Symposium on Security and Privacy (SP)}}. IEEE, \bibinfo{pages}{1071--1087}.
\newblock


\bibitem[Hedin et~al\mbox{.}(2014)]%
        {hedin2014jsflow}
\bibfield{author}{\bibinfo{person}{Daniel Hedin}, \bibinfo{person}{Arnar
  Birgisson}, \bibinfo{person}{Luciano Bello}, {and} \bibinfo{person}{Andrei
  Sabelfeld}.} \bibinfo{year}{2014}\natexlab{}.
\newblock \showarticletitle{JSFlow: Tracking information flow in JavaScript and
  its APIs}. In \bibinfo{booktitle}{\emph{Proceedings of the 29th Annual ACM
  Symposium on Applied Computing}}. \bibinfo{pages}{1663--1671}.
\newblock


\bibitem[Hedin and Sabelfeld(2012)]%
        {hedin2012information}
\bibfield{author}{\bibinfo{person}{Daniel Hedin} {and} \bibinfo{person}{Andrei
  Sabelfeld}.} \bibinfo{year}{2012}\natexlab{}.
\newblock \showarticletitle{Information-flow security for a core of
  JavaScript}. In \bibinfo{booktitle}{\emph{2012 IEEE 25th Computer Security
  Foundations Symposium}}. IEEE, \bibinfo{pages}{3--18}.
\newblock


\bibitem[Huang et~al\mbox{.}(2015)]%
        {huang2015supor}
\bibfield{author}{\bibinfo{person}{Jianjun Huang}, \bibinfo{person}{Zhichun
  Li}, \bibinfo{person}{Xusheng Xiao}, \bibinfo{person}{Zhenyu Wu},
  \bibinfo{person}{Kangjie Lu}, \bibinfo{person}{Xiangyu Zhang}, {and}
  \bibinfo{person}{Guofei Jiang}.} \bibinfo{year}{2015}\natexlab{}.
\newblock \showarticletitle{$\{$SUPOR$\}$: Precise and scalable sensitive user
  input detection for android apps}. In \bibinfo{booktitle}{\emph{24th USENIX
  Security Symposium (USENIX Security 15)}}. \bibinfo{pages}{977--992}.
\newblock


\bibitem[Huang et~al\mbox{.}(2016)]%
        {huang2016detecting}
\bibfield{author}{\bibinfo{person}{Jianjun Huang}, \bibinfo{person}{Xiangyu
  Zhang}, {and} \bibinfo{person}{Lin Tan}.} \bibinfo{year}{2016}\natexlab{}.
\newblock \showarticletitle{Detecting sensitive data disclosure via
  bi-directional text correlation analysis}. In
  \bibinfo{booktitle}{\emph{Proceedings of the 2016 24th ACM SIGSOFT
  International Symposium on Foundations of Software Engineering}}.
  \bibinfo{pages}{169--180}.
\newblock


\bibitem[Huang et~al\mbox{.}(2014)]%
        {huang2014asdroid}
\bibfield{author}{\bibinfo{person}{Jianjun Huang}, \bibinfo{person}{Xiangyu
  Zhang}, \bibinfo{person}{Lin Tan}, \bibinfo{person}{Peng Wang}, {and}
  \bibinfo{person}{Bin Liang}.} \bibinfo{year}{2014}\natexlab{}.
\newblock \showarticletitle{Asdroid: Detecting stealthy behaviors in android
  applications by user interface and program behavior contradiction}. In
  \bibinfo{booktitle}{\emph{Proceedings of the 36th International Conference on
  Software Engineering}}. \bibinfo{pages}{1036--1046}.
\newblock


\bibitem[Jensen et~al\mbox{.}(2011)]%
        {jensen2011modeling}
\bibfield{author}{\bibinfo{person}{Simon~Holm Jensen}, \bibinfo{person}{Magnus
  Madsen}, {and} \bibinfo{person}{Anders M{\o}ller}.}
  \bibinfo{year}{2011}\natexlab{}.
\newblock \showarticletitle{Modeling the HTML DOM and browser API in static
  analysis of JavaScript web applications}. In
  \bibinfo{booktitle}{\emph{Proceedings of the 19th ACM SIGSOFT symposium and
  the 13th European conference on Foundations of software engineering}}.
  \bibinfo{pages}{59--69}.
\newblock


\bibitem[Jin et~al\mbox{.}(2018)]%
        {jin2018they}
\bibfield{author}{\bibinfo{person}{Haojian Jin}, \bibinfo{person}{Minyi Liu},
  \bibinfo{person}{Kevan Dodhia}, \bibinfo{person}{Yuanchun Li},
  \bibinfo{person}{Gaurav Srivastava}, \bibinfo{person}{Matthew Fredrikson},
  \bibinfo{person}{Yuvraj Agarwal}, {and} \bibinfo{person}{Jason~I Hong}.}
  \bibinfo{year}{2018}\natexlab{}.
\newblock \showarticletitle{Why are they collecting my data? inferring the
  purposes of network traffic in mobile apps}.
\newblock \bibinfo{journal}{\emph{Proceedings of the ACM on Interactive,
  Mobile, Wearable and Ubiquitous Technologies}} \bibinfo{volume}{2},
  \bibinfo{number}{4} (\bibinfo{year}{2018}), \bibinfo{pages}{1--27}.
\newblock


\bibitem[Kashyap et~al\mbox{.}(2014)]%
        {kashyap2014jsai}
\bibfield{author}{\bibinfo{person}{Vineeth Kashyap}, \bibinfo{person}{Kyle
  Dewey}, \bibinfo{person}{Ethan~A Kuefner}, \bibinfo{person}{John Wagner},
  \bibinfo{person}{Kevin Gibbons}, \bibinfo{person}{John Sarracino},
  \bibinfo{person}{Ben Wiedermann}, {and} \bibinfo{person}{Ben Hardekopf}.}
  \bibinfo{year}{2014}\natexlab{}.
\newblock \showarticletitle{JSAI: A static analysis platform for JavaScript}.
  In \bibinfo{booktitle}{\emph{Proceedings of the 22nd ACM SIGSOFT
  international symposium on Foundations of Software Engineering}}.
  \bibinfo{pages}{121--132}.
\newblock


\bibitem[Li et~al\mbox{.}(2015)]%
        {li2015iccta}
\bibfield{author}{\bibinfo{person}{Li Li}, \bibinfo{person}{Alexandre Bartel},
  \bibinfo{person}{Tegawend{\'e}~F Bissyand{\'e}}, \bibinfo{person}{Jacques
  Klein}, \bibinfo{person}{Yves Le~Traon}, \bibinfo{person}{Steven Arzt},
  \bibinfo{person}{Siegfried Rasthofer}, \bibinfo{person}{Eric Bodden},
  \bibinfo{person}{Damien Octeau}, {and} \bibinfo{person}{Patrick McDaniel}.}
  \bibinfo{year}{2015}\natexlab{}.
\newblock \showarticletitle{Iccta: Detecting inter-component privacy leaks in
  android apps}. In \bibinfo{booktitle}{\emph{2015 IEEE/ACM 37th IEEE
  International Conference on Software Engineering}}, Vol.~\bibinfo{volume}{1}.
  IEEE, \bibinfo{pages}{280--291}.
\newblock


\bibitem[Li et~al\mbox{.}(2017)]%
        {li2017static}
\bibfield{author}{\bibinfo{person}{Li Li}, \bibinfo{person}{Tegawend{\'e}~F
  Bissyand{\'e}}, \bibinfo{person}{Mike Papadakis}, \bibinfo{person}{Siegfried
  Rasthofer}, \bibinfo{person}{Alexandre Bartel}, \bibinfo{person}{Damien
  Octeau}, \bibinfo{person}{Jacques Klein}, {and} \bibinfo{person}{Le Traon}.}
  \bibinfo{year}{2017}\natexlab{}.
\newblock \showarticletitle{Static analysis of android apps: A systematic
  literature review}.
\newblock \bibinfo{journal}{\emph{Information and Software Technology}}
  \bibinfo{volume}{88} (\bibinfo{year}{2017}), \bibinfo{pages}{67--95}.
\newblock


\bibitem[Liu et~al\mbox{.}(2020)]%
        {liu2020industry}
\bibfield{author}{\bibinfo{person}{Yi Liu}, \bibinfo{person}{Jinhui Xie},
  \bibinfo{person}{Jianbo Yang}, \bibinfo{person}{Shiyu Guo},
  \bibinfo{person}{Yuetang Deng}, \bibinfo{person}{Shuqing Li},
  \bibinfo{person}{Yechang Wu}, {and} \bibinfo{person}{Yepang Liu}.}
  \bibinfo{year}{2020}\natexlab{}.
\newblock \showarticletitle{Industry practice of Javascript dynamic analysis on
  WeChat mini-programs}. In \bibinfo{booktitle}{\emph{2020 35th IEEE/ACM
  International Conference on Automated Software Engineering (ASE)}}. IEEE,
  \bibinfo{pages}{1189--1193}.
\newblock


\bibitem[Lu et~al\mbox{.}(2020)]%
        {lu2020demystifying}
\bibfield{author}{\bibinfo{person}{Haoran Lu}, \bibinfo{person}{Luyi Xing},
  \bibinfo{person}{Yue Xiao}, \bibinfo{person}{Yifan Zhang},
  \bibinfo{person}{Xiaojing Liao}, \bibinfo{person}{XiaoFeng Wang}, {and}
  \bibinfo{person}{Xueqiang Wang}.} \bibinfo{year}{2020}\natexlab{}.
\newblock \showarticletitle{Demystifying resource management risks in emerging
  mobile app-in-app ecosystems}. In \bibinfo{booktitle}{\emph{Proceedings of
  the 2020 ACM SIGSAC Conference on Computer and Communications Security}}.
  \bibinfo{pages}{569--585}.
\newblock


\bibitem[Ma(2019a)]%
        {ma2019app}
\bibfield{author}{\bibinfo{person}{Xiaojuan Ma}.}
  \bibinfo{year}{2019}\natexlab{a}.
\newblock \showarticletitle{App store Killer? The storm of WeChat mini programs
  swept over the mobile app ecosystem}.
\newblock \bibinfo{journal}{\emph{Retrieved November}}  \bibinfo{volume}{15}
  (\bibinfo{year}{2019}), \bibinfo{pages}{2019}.
\newblock


\bibitem[Ma(2019b)]%
        {ma2019appstorekiller}
\bibfield{author}{\bibinfo{person}{Xiaojuan Ma}.}
  \bibinfo{year}{2019}\natexlab{b}.
\newblock \showarticletitle{App store Killer? The storm of WeChat mini programs
  swept over the mobile app ecosystem}.
\newblock \bibinfo{journal}{\emph{Retrieved November}}  \bibinfo{volume}{15}
  (\bibinfo{year}{2019}), \bibinfo{pages}{2019}.
\newblock


\bibitem[Mikolov et~al\mbox{.}(2013)]%
        {mikolov2013word2vec}
\bibfield{author}{\bibinfo{person}{Tomas Mikolov}, \bibinfo{person}{Kai Chen},
  \bibinfo{person}{Greg Corrado}, {and} \bibinfo{person}{Jeffrey Dean}.}
  \bibinfo{year}{2013}\natexlab{}.
\newblock \showarticletitle{Efficient estimation of word representations in
  vector space}.
\newblock \bibinfo{journal}{\emph{arXiv preprint arXiv:1301.3781}}
  (\bibinfo{year}{2013}).
\newblock


\bibitem[Nan et~al\mbox{.}(2015)]%
        {nan2015uipicker}
\bibfield{author}{\bibinfo{person}{Yuhong Nan}, \bibinfo{person}{Min Yang},
  \bibinfo{person}{Zhemin Yang}, \bibinfo{person}{Shunfan Zhou},
  \bibinfo{person}{Guofei Gu}, {and} \bibinfo{person}{XiaoFeng Wang}.}
  \bibinfo{year}{2015}\natexlab{}.
\newblock \showarticletitle{$\{$UIPicker$\}$:$\{$User-Input$\}$ Privacy
  Identification in Mobile Applications}. In \bibinfo{booktitle}{\emph{24th
  USENIX Security Symposium (USENIX Security 15)}}. \bibinfo{pages}{993--1008}.
\newblock


\bibitem[Nishant Das~Patnaik(2022)]%
        {JsPrime}
\bibfield{author}{\bibinfo{person}{Sarathi Sabyasachi~Sahoo Nishant
  Das~Patnaik}.} \bibinfo{year}{2022}\natexlab{}.
\newblock \bibinfo{title}{JSPrime: a javascript static security analysis tool}.
\newblock \bibinfo{howpublished}{\url{https://dpnishant.github.io/jsprime/}}.
\newblock
\newblock
\shownote{[Online; accessed 1-January-2022]}.


\bibitem[Octeau et~al\mbox{.}(2013)]%
        {octeau2013effective}
\bibfield{author}{\bibinfo{person}{Damien Octeau}, \bibinfo{person}{Patrick
  McDaniel}, \bibinfo{person}{Somesh Jha}, \bibinfo{person}{Alexandre Bartel},
  \bibinfo{person}{Eric Bodden}, \bibinfo{person}{Jacques Klein}, {and}
  \bibinfo{person}{Yves Le~Traon}.} \bibinfo{year}{2013}\natexlab{}.
\newblock \showarticletitle{Effective $\{$Inter-Component$\}$ Communication
  Mapping in Android: An Essential Step Towards Holistic Security Analysis}. In
  \bibinfo{booktitle}{\emph{22nd USENIX Security Symposium (USENIX Security
  13)}}. \bibinfo{pages}{543--558}.
\newblock


\bibitem[Pan et~al\mbox{.}(2018)]%
        {pan2018flowcog}
\bibfield{author}{\bibinfo{person}{Xiang Pan}, \bibinfo{person}{Yinzhi Cao},
  \bibinfo{person}{Xuechao Du}, \bibinfo{person}{Boyuan He},
  \bibinfo{person}{Gan Fang}, \bibinfo{person}{Rui Shao}, {and}
  \bibinfo{person}{Yan Chen}.} \bibinfo{year}{2018}\natexlab{}.
\newblock \showarticletitle{$\{$FlowCog$\}$: Context-aware Semantics Extraction
  and Analysis of Information Flow Leaks in Android Apps}. In
  \bibinfo{booktitle}{\emph{27th USENIX Security Symposium (USENIX Security
  18)}}. \bibinfo{pages}{1669--1685}.
\newblock


\bibitem[Park et~al\mbox{.}(2017)]%
        {park2017analysis}
\bibfield{author}{\bibinfo{person}{Jihyeok Park}, \bibinfo{person}{Yeonhee
  Ryou}, \bibinfo{person}{Joonyoung Park}, {and} \bibinfo{person}{Sukyoung
  Ryu}.} \bibinfo{year}{2017}\natexlab{}.
\newblock \showarticletitle{Analysis of JavaScript web applications using SAFE
  2.0}. In \bibinfo{booktitle}{\emph{2017 IEEE/ACM 39th International
  Conference on Software Engineering Companion (ICSE-C)}}. IEEE,
  \bibinfo{pages}{59--62}.
\newblock


\bibitem[Qu et~al\mbox{.}(2014)]%
        {qu2014autocog}
\bibfield{author}{\bibinfo{person}{Zhengyang Qu}, \bibinfo{person}{Vaibhav
  Rastogi}, \bibinfo{person}{Xinyi Zhang}, \bibinfo{person}{Yan Chen},
  \bibinfo{person}{Tiantian Zhu}, {and} \bibinfo{person}{Zhong Chen}.}
  \bibinfo{year}{2014}\natexlab{}.
\newblock \showarticletitle{Autocog: Measuring the description-to-permission
  fidelity in android applications}. In \bibinfo{booktitle}{\emph{Proceedings
  of the 2014 ACM SIGSAC Conference on Computer and Communications Security}}.
  \bibinfo{pages}{1354--1365}.
\newblock


\bibitem[Rahman et~al\mbox{.}(2022)]%
        {rahman2022permpress}
\bibfield{author}{\bibinfo{person}{Muhammad~Sajidur Rahman},
  \bibinfo{person}{Pirouz Naghavi}, \bibinfo{person}{Blas Kojusner},
  \bibinfo{person}{Sadia Afroz}, \bibinfo{person}{Byron Williams},
  \bibinfo{person}{Sara Rampazzi}, {and} \bibinfo{person}{Vincent
  Bindschaedler}.} \bibinfo{year}{2022}\natexlab{}.
\newblock \showarticletitle{PermPress: Machine Learning-Based Pipeline to
  Evaluate Permissions in App Privacy Policies}.
\newblock \bibinfo{journal}{\emph{IEEE Access}}  \bibinfo{volume}{10}
  (\bibinfo{year}{2022}), \bibinfo{pages}{89248--89269}.
\newblock


\bibitem[Slavin et~al\mbox{.}(2016)]%
        {slavin2016toward}
\bibfield{author}{\bibinfo{person}{Rocky Slavin}, \bibinfo{person}{Xiaoyin
  Wang}, \bibinfo{person}{Mitra~Bokaei Hosseini}, \bibinfo{person}{James
  Hester}, \bibinfo{person}{Ram Krishnan}, \bibinfo{person}{Jaspreet Bhatia},
  \bibinfo{person}{Travis~D Breaux}, {and} \bibinfo{person}{Jianwei Niu}.}
  \bibinfo{year}{2016}\natexlab{}.
\newblock \showarticletitle{Toward a framework for detecting privacy policy
  violations in android application code}. In
  \bibinfo{booktitle}{\emph{Proceedings of the 38th International Conference on
  Software Engineering}}. \bibinfo{pages}{25--36}.
\newblock


\bibitem[Story et~al\mbox{.}(2019)]%
        {story2019compliance}
\bibfield{author}{\bibinfo{person}{Peter Story}, \bibinfo{person}{Sebastian
  Zimmeck}, \bibinfo{person}{Abhilasha Ravichander}, \bibinfo{person}{Daniel
  Smullen}, \bibinfo{person}{Ziqi Wang}, \bibinfo{person}{Joel Reidenberg},
  \bibinfo{person}{N~Cameron Russell}, {and} \bibinfo{person}{Norman Sadeh}.}
  \bibinfo{year}{2019}\natexlab{}.
\newblock \showarticletitle{Natural language processing for mobile app privacy
  compliance}. In \bibinfo{booktitle}{\emph{AAAI Spring Symposium on
  Privacy-Enhancing Artificial Intelligence and Language Technologies}},
  Vol.~\bibinfo{volume}{10}.
\newblock


\bibitem[Sunyaev et~al\mbox{.}(2015)]%
        {sunyaev2015policystatistic}
\bibfield{author}{\bibinfo{person}{Ali Sunyaev}, \bibinfo{person}{Tobias
  Dehling}, \bibinfo{person}{Patrick~L Taylor}, {and}
  \bibinfo{person}{Kenneth~D Mandl}.} \bibinfo{year}{2015}\natexlab{}.
\newblock \showarticletitle{Availability and quality of mobile health app
  privacy policies}.
\newblock \bibinfo{journal}{\emph{Journal of the American Medical Informatics
  Association}} \bibinfo{volume}{22}, \bibinfo{number}{e1}
  (\bibinfo{year}{2015}), \bibinfo{pages}{e28--e33}.
\newblock


\bibitem[Tesfay et~al\mbox{.}(2018)]%
        {tesfay2018privacyguide}
\bibfield{author}{\bibinfo{person}{Welderufael~B Tesfay},
  \bibinfo{person}{Peter Hofmann}, \bibinfo{person}{Toru Nakamura},
  \bibinfo{person}{Shinsaku Kiyomoto}, {and} \bibinfo{person}{Jetzabel Serna}.}
  \bibinfo{year}{2018}\natexlab{}.
\newblock \showarticletitle{PrivacyGuide: towards an implementation of the EU
  GDPR on internet privacy policy evaluation}. In
  \bibinfo{booktitle}{\emph{Proceedings of the Fourth ACM International
  Workshop on Security and Privacy Analytics}}. \bibinfo{pages}{15--21}.
\newblock


\bibitem[Wang et~al\mbox{.}(2022)]%
        {wang2022characterizing}
\bibfield{author}{\bibinfo{person}{Tao Wang}, \bibinfo{person}{Qingxin Xu},
  \bibinfo{person}{Xiaoning Chang}, \bibinfo{person}{Wensheng Dou},
  \bibinfo{person}{Jiaxin Zhu}, \bibinfo{person}{Jinhui Xie},
  \bibinfo{person}{Yuetang Deng}, \bibinfo{person}{Jianbo Yang},
  \bibinfo{person}{Jiaheng Yang}, \bibinfo{person}{Jun Wei}, {et~al\mbox{.}}}
  \bibinfo{year}{2022}\natexlab{}.
\newblock \showarticletitle{Characterizing and detecting bugs in WeChat
  mini-programs}. In \bibinfo{booktitle}{\emph{Proceedings of the 44th
  International Conference on Software Engineering}}.
  \bibinfo{pages}{363--375}.
\newblock


\bibitem[Wang et~al\mbox{.}(2018)]%
        {wang2018guileak}
\bibfield{author}{\bibinfo{person}{Xiaoyin Wang}, \bibinfo{person}{Xue Qin},
  \bibinfo{person}{Mitra~Bokaei Hosseini}, \bibinfo{person}{Rocky Slavin},
  \bibinfo{person}{Travis~D Breaux}, {and} \bibinfo{person}{Jianwei Niu}.}
  \bibinfo{year}{2018}\natexlab{}.
\newblock \showarticletitle{Guileak: Tracing privacy policy claims on user
  input data for android applications}. In
  \bibinfo{booktitle}{\emph{Proceedings of the 40th International Conference on
  Software Engineering}}. \bibinfo{pages}{37--47}.
\newblock


\bibitem[Wei et~al\mbox{.}(2018)]%
        {wei2018amandroid}
\bibfield{author}{\bibinfo{person}{Fengguo Wei}, \bibinfo{person}{Sankardas
  Roy}, {and} \bibinfo{person}{Xinming Ou}.} \bibinfo{year}{2018}\natexlab{}.
\newblock \showarticletitle{Amandroid: A precise and general inter-component
  data flow analysis framework for security vetting of android apps}.
\newblock \bibinfo{journal}{\emph{ACM Transactions on Privacy and Security
  (TOPS)}} \bibinfo{volume}{21}, \bibinfo{number}{3} (\bibinfo{year}{2018}),
  \bibinfo{pages}{1--32}.
\newblock


\bibitem[Xiao et~al\mbox{.}(2019)]%
        {xiao2019iconintent}
\bibfield{author}{\bibinfo{person}{Xusheng Xiao}, \bibinfo{person}{Xiaoyin
  Wang}, \bibinfo{person}{Zhihao Cao}, \bibinfo{person}{Hanlin Wang}, {and}
  \bibinfo{person}{Peng Gao}.} \bibinfo{year}{2019}\natexlab{}.
\newblock \showarticletitle{Iconintent: automatic identification of sensitive
  ui widgets based on icon classification for android apps}. In
  \bibinfo{booktitle}{\emph{2019 IEEE/ACM 41st International Conference on
  Software Engineering (ICSE)}}. IEEE, \bibinfo{pages}{257--268}.
\newblock


\bibitem[Xue et~al\mbox{.}(2018)]%
        {xue2018ndroid}
\bibfield{author}{\bibinfo{person}{Lei Xue}, \bibinfo{person}{Chenxiong Qian},
  \bibinfo{person}{Hao Zhou}, \bibinfo{person}{Xiapu Luo},
  \bibinfo{person}{Yajin Zhou}, \bibinfo{person}{Yuru Shao}, {and}
  \bibinfo{person}{Alvin~TS Chan}.} \bibinfo{year}{2018}\natexlab{}.
\newblock \showarticletitle{NDroid: Toward tracking information flows across
  multiple Android contexts}.
\newblock \bibinfo{journal}{\emph{IEEE Transactions on Information Forensics
  and Security}} \bibinfo{volume}{14}, \bibinfo{number}{3}
  (\bibinfo{year}{2018}), \bibinfo{pages}{814--828}.
\newblock


\bibitem[Yu et~al\mbox{.}(2018)]%
        {yu2018ppchecker}
\bibfield{author}{\bibinfo{person}{Le Yu}, \bibinfo{person}{Xiapu Luo},
  \bibinfo{person}{Jiachi Chen}, \bibinfo{person}{Hao Zhou},
  \bibinfo{person}{Tao Zhang}, \bibinfo{person}{Henry Chang}, {and}
  \bibinfo{person}{Hareton~KN Leung}.} \bibinfo{year}{2018}\natexlab{}.
\newblock \showarticletitle{Ppchecker: Towards accessing the trustworthiness of
  android apps’ privacy policies}.
\newblock \bibinfo{journal}{\emph{IEEE Transactions on Software Engineering}}
  \bibinfo{volume}{47}, \bibinfo{number}{2} (\bibinfo{year}{2018}),
  \bibinfo{pages}{221--242}.
\newblock


\bibitem[Yu et~al\mbox{.}(2016)]%
        {yu2016can}
\bibfield{author}{\bibinfo{person}{Le Yu}, \bibinfo{person}{Xiapu Luo},
  \bibinfo{person}{Xule Liu}, {and} \bibinfo{person}{Tao Zhang}.}
  \bibinfo{year}{2016}\natexlab{}.
\newblock \showarticletitle{Can we trust the privacy policies of android
  apps?}. In \bibinfo{booktitle}{\emph{2016 46th Annual IEEE/IFIP International
  Conference on Dependable Systems and Networks (DSN)}}. IEEE,
  \bibinfo{pages}{538--549}.
\newblock


\bibitem[Zhang et~al\mbox{.}(2021)]%
        {zhang2021measurement}
\bibfield{author}{\bibinfo{person}{Yue Zhang}, \bibinfo{person}{Bayan
  Turkistani}, \bibinfo{person}{Allen~Yuqing Yang}, \bibinfo{person}{Chaoshun
  Zuo}, {and} \bibinfo{person}{Zhiqiang Lin}.} \bibinfo{year}{2021}\natexlab{}.
\newblock \showarticletitle{A Measurement Study of Wechat Mini-Apps}. In
  \bibinfo{booktitle}{\emph{Abstract Proceedings of the 2021 ACM
  SIGMETRICS/International Conference on Measurement and Modeling of Computer
  Systems}}. \bibinfo{pages}{19--20}.
\newblock


\bibitem[Zimmeck et~al\mbox{.}(2019)]%
        {zimmeck2019maps}
\bibfield{author}{\bibinfo{person}{Sebastian Zimmeck}, \bibinfo{person}{Peter
  Story}, \bibinfo{person}{Daniel Smullen}, \bibinfo{person}{Abhilasha
  Ravichander}, \bibinfo{person}{Ziqi Wang}, \bibinfo{person}{Joel~R
  Reidenberg}, \bibinfo{person}{N~Cameron Russell}, {and}
  \bibinfo{person}{Norman Sadeh}.} \bibinfo{year}{2019}\natexlab{}.
\newblock \showarticletitle{Maps: Scaling privacy compliance analysis to a
  million apps}.
\newblock \bibinfo{journal}{\emph{Proc. Priv. Enhancing Tech.}}
  \bibinfo{volume}{2019} (\bibinfo{year}{2019}), \bibinfo{pages}{66}.
\newblock


\bibitem[Zimmeck et~al\mbox{.}(2016)]%
        {zimmeck2016automated}
\bibfield{author}{\bibinfo{person}{Sebastian Zimmeck}, \bibinfo{person}{Ziqi
  Wang}, \bibinfo{person}{Lieyong Zou}, \bibinfo{person}{Roger Iyengar},
  \bibinfo{person}{Bin Liu}, \bibinfo{person}{Florian Schaub},
  \bibinfo{person}{Shomir Wilson}, \bibinfo{person}{Norman Sadeh},
  \bibinfo{person}{Steven Bellovin}, {and} \bibinfo{person}{Joel Reidenberg}.}
  \bibinfo{year}{2016}\natexlab{}.
\newblock \showarticletitle{Automated analysis of privacy requirements for
  mobile apps}. In \bibinfo{booktitle}{\emph{2016 AAAI Fall Symposium Series}}.
\newblock


\end{thebibliography}

\end{document}